\PassOptionsToPackage{dvipsnames,svgnames,table}{xcolor}

\documentclass[nohyperref]{article}

\usepackage{microtype}
\usepackage{graphicx}
\usepackage{booktabs} %

\usepackage{hyperref}

\usepackage[accepted]{icml2023}

\usepackage{amsmath}
\usepackage{amssymb}
\usepackage{mathtools}
\usepackage{amsthm}

\usepackage[capitalize,noabbrev]{cleveref}

\theoremstyle{plain}

\theoremstyle{definition}

\theoremstyle{remark}

\usepackage[textsize=tiny]{todonotes}

\usepackage{amsfonts}
\usepackage[utf8]{inputenc}
\usepackage{multirow}
\usepackage{makecell}
\usepackage{courier}
\usepackage{xspace}
\usepackage{bm}
\usepackage{wrapfig}
\usepackage{oubraces}
\usepackage{amsthm}
\usepackage{caption}
\usepackage{subcaption}
\usepackage{stfloats}
\usepackage{url}
\usepackage{enumitem}

\definecolor{Gray}{gray}{0.7}
\newcolumntype{g}{>{\columncolor{Gray}}r}

\DeclareMathOperator{\R}{\mathbb{R}}

\newcommand{\norm}[1]{\left\lVert#1\right\rVert}

\DeclareMathOperator*{\argmin}{arg\,min}

\def\eqref#1{Eqn.~(\ref{#1})}

\newcommand{\bp}{Bypassing\xspace}
\newcommand{\nbp}{Biased-Gradient\xspace}

\renewcommand{\algorithmiccomment}[1]{{\color{gray}\small \# \emph{#1}}}

\icmltitlerunning{Realistic Decision-Based Attacks on Machine Learning Systems\hfill\thepage}

\begin{document}

\twocolumn[
    \icmltitle{\resizebox{0.98\textwidth}{!}{Preprocessors Matter! Realistic Decision-Based Attacks on Machine Learning Systems}}

    \icmlsetsymbol{equal}{*}

    \begin{icmlauthorlist}
        \icmlauthor{Chawin Sitawarin}{ucb}
        \icmlauthor{Florian Tram\`er}{eth}
        \icmlauthor{Nicholas Carlini}{google}
    \end{icmlauthorlist}

    \icmlaffiliation{ucb}{Department of Computer Science, University of California, Berkeley, USA. Work partially done while the author was at Google.}
    \icmlaffiliation{google}{Google DeepMind, Mountain View, USA}
    \icmlaffiliation{eth}{ETH Z\"urich, Z\"urich, Switzerland.}

    \icmlcorrespondingauthor{Chawin Sitawarin}{chawins@berkeley.edu}

    \icmlkeywords{Machine Learning, ICML, Adversarial Example, Black-Box Attack, Decision-Based Attack, Preprocessors, Extraction Attack}

    \vskip 0.3in
]

\printAffiliationsAndNotice{}  %

\begin{abstract}
    Decision-based attacks construct adversarial examples
    against a machine learning (ML) model by making only hard-label queries.
    These attacks have mainly been applied directly
    to standalone neural networks.
    However, in practice, ML models are just one component of a larger learning system.
    We find that by adding a \emph{single} preprocessor in front of a classifier, state-of-the-art query-based attacks are up to \emph{seven}$\times$ less effective at attacking a prediction pipeline than at attacking the model alone.
    We explain this discrepancy by the fact that most preprocessors introduce some notion of \emph{invariance} to the input space.
    Hence, attacks that are unaware of this invariance inevitably waste a large number of queries to re-discover or overcome it.
    We therefore develop techniques to (i) reverse-engineer the preprocessor
    and then (ii) use this extracted information to attack the end-to-end system.
    Our preprocessors extraction method requires only a few hundreds queries, and our preprocessor-aware attacks recover the same efficacy as when attacking the model alone.
    The code can be found at \url{https://github.com/google-research/preprocessor-aware-black-box-attack}.
\end{abstract}

\section{Introduction}

Machine learning is widely used in security-critical systems, for example for detecting abusive, harmful or otherwise unsafe online content~\citep{waseem-etal-2017-understanding,clarifai_best_,jha-mamidi-2017-compliment}.
It is critical that such systems are robust against adversaries who seeks to evade them.

Yet, an extensive body of work has shown that an adversary can fool machine learning models with \emph{adversarial examples}~\citep{biggio_evasion_2013,szegedy_intriguing_2014}.
Most prior work focuses on \emph{white-box} attacks, 
where an adversary has perfect knowledge of the entire machine
learning system~\citep{carlini_evaluating_2017}.
Yet, real adversaries rarely have this level of access~\citep{tramer_adversarial_2019a},
and must thus instead resort to \emph{black-box} attacks~\citep{chen_zoo_2017}.
\emph{Decision-based} attacks~\citep{brendel_decisionbased_2018} are a particularly practical attack vector, as these attacks only require the ability to query a target model and observe its decisions.

However, existing decision-based attacks~\citep{brendel_adversarial_2018,cheng_signopt_2020,chen_hopskipjumpattack_2020,li_qeba_2020} have primarily been evaluated against standalone ML models ``\emph{in the lab}'', thereby ignoring the components of broader learning systems that are used in practice.
While some decision-based attacks have been demonstrated on production systems as a proof-of-concept (e.g., \citet{ilyas_blackbox_2018, brendel_decisionbased_2018, li_qeba_2020}), it is not well understood how these attacks perform on end-to-end learning systems compared to standalone models.

We show that \textbf{existing decision-based attacks are significantly less effective against end-to-end systems compared to standalone machine learning models.}
For example, a standard decision-based attack can evade a ResNet image classifier on ImageNet with an average $\ell_2$-distortion of $3.7$ (defined formally later).
Yet, if we instead attack an end-to-end learning system that simply \emph{preprocesses} the classifier's input before classifying it---e.g., by resizing or compressing the image---the attack achieves an average $\ell_2$ distortion of $28.5$---\textbf{a $\mathbf{7\times}$ increase!}
We further find that extensive hyperparameter tuning and running the attacks for more iterations fail to resolve this issue.
We thus argue that existing decision-box attacks have fundamental limitations
that make them sub-optimal in practice.

To remedy this, \textbf{we develop improved attacks that achieve the same success rate when attacking systems with unknown preprocessors, as when attacking standalone models}.
Our attacks combine decision-based attacks with techniques developed for model extraction~\citep{tramer_stealing_2016}.
Our attacks first query the system to reverse-engineer
the preprocessor(s) used in the input pipeline,
and then mount a modified \emph{preprocessor-aware} decision-based attack.
Our extraction procedure is efficient and often requires only a few hundred
queries to identify commonly used preprocessors. This cost can also be amortized across many generated adversarial examples.
We find that even the \emph{least efficient} preprocessor-aware attack outperforms \emph{all} unaware attacks.
Learning the system's preprocessing pipeline is thus more important than devising an efficient standalone attack.

\section{Background and Related Work}

\textbf{Adversarial Examples.}
Adversarial examples are inputs designed to fool a machine learning classifier~\citep{biggio_evasion_2013,szegedy_intriguing_2014,goodfellow_explaining_2015}.
For some classifier $f$, an example $x$ has an
adversarial example $x' = x + \delta$ if
$f(x) \ne f(x')$, where $\delta$ is a small perturbation under some $\ell_p$-norm, i.e., $\norm{\delta}_p \le \epsilon$.
Adversarial examples can be constructed either in the white-box setting (where the adversary uses gradient descent to produce the perturbation $\delta$)~\citep{carlini_evaluating_2017,madry_deep_2018},
or more realistically, in the black-box setting (where the adversary uses just query access to the system)~\citep{papernot_practical_2017,chen_zoo_2017,brendel_decisionbased_2018}.
Our paper focuses on this black-box setting with $\ell_2$-norm perturbations.

\emph{Decision-based} can generate adversarial examples
with only query access to the remote model's decisions (i.e., the output class $y \gets f(x)$).
These attacks typically work by finding the decision boundary between the original image and a target label of interest and then walking along the decision boundary to reduce the total distortion~\citep{brendel_decisionbased_2018,cheng_signopt_2020,chen_hopskipjumpattack_2020,li_qeba_2020}.

It has been shown that decision-based attacks should
operate at the lowest-dimensional input space possible.
For example, QEBA~\citep{li_qeba_2020} improves upon HSJA~\cite{chen_hopskipjumpattack_2020} by constructing adversarial examples in a lower-dimensional embedding space.
This phenomenon will help explain some of the results we observe,
where we find that high-dimensional images require more queries to attack.

Adversarial examples need not exploit the classifier itself.
\emph{Image scaling attacks}~\citep{quiring_adversarial_2020} construct a high-resolution image $x$ so that after resizing to a smaller $\hat x$, the low resolution image is visually dissimilar to $x$.
As a result, any accurate classifier will (correctly) classify the high-resolution image and the low-resolution image differently.
\citet{gao_rethinking_2022} consider the image-scaling attack in conjunction with a classifier similar to our setting.
However, our work applies to arbitrary preprocessors, not limited to resizing, and we also propose an extraction attack to unveil the deployed preprocessor in the first place.

\textbf{Preprocessing defenses.}
A number of proposed defenses against adversarial examples preprocess inputs before classification~\citep{guo_countering_2018,song_pixeldefend_2018}.
Unfortunately, these defenses are largely ineffective in a white-box setting~\citep{athalye_obfuscated_2018,tramer_adaptive_2020,sitawarin_demystifying_2022}.
Surprisingly, recent work has shown that defending against existing decision-based attacks with preprocessors is quite simple.
\citet{aithal2022mitigating,qin2021random} show that adding small amounts of random noise to inputs impedes all current attacks.
This suggests that there may be a significant gap between
the capabilities of white-box and black-box attacks when preprocessors are present.

\textbf{Model Stealing Attacks.}
To improve the efficacy of black-box attacks, we make use of techniques from model stealing attacks~\citep{tramer_stealing_2016}.
These attacks aim to create a ML model that closely mimics the behavior of a remote model~\citep{jagielski_high_2020}.
Our goal is slightly different as we only aim to ``steal'' the system's preprocessor and use this knowledge to mount stronger evasion attacks.
For this, we leverage techniques that have been used to extract \emph{functionally equivalent} models, which exactly match the behavior of the remote model on all inputs~\citep{milli2019model,rolnick2020reverse,carlini2020cryptanalytic}.

\section{Setup and Threat Model} \label{sec:setup}

\subsection{Notation}

We denote an unperturbed input image in the \textit{original space} as $x_o \in \mathcal{X}_o \coloneqq [0,1]^{s_o \times s_o}$ and a processed image in the \textit{model space} as $x_m \in \mathcal{X}_m \subseteq [0,1]^{s_m \times s_m}$.
The original size $s_o$ can be the same or different from the target size $s_m$.
A preprocessor $t: \mathcal{X}_o \to \mathcal{X}_m$ maps $x_o$ to $x_m \coloneqq t(x_o)$.
For instance, a resizing preprocessor that maps an image of size $256 \times 256$ pixels to $224 \times 224$ pixels means that $s_o=256$, $s_m=224$, and $\mathcal{X}_m = [0,1]^{224 \times 224}$.
As another example, an 8-bit quantization restricts $\mathcal{X}_m$ to a discrete space of $\{0, 1/255, 2/255, \dots, 1\}^{s_m \times s_m}$ and $s_o = s_m$.
The classifier, excluding the preprocessor, is represented by $f:\mathcal{X}_m \to \mathcal{Y}$ where $\mathcal{Y}$ is the hard label space.
Finally, the entire classification pipeline is denoted by $f \circ t:\mathcal{X}_o \to \mathcal{Y}$.

\subsection{Threat Model} \label{ssec:threat_model}

The key distinguishing factor between previous works and ours is that we consider a \textbf{preprocessing pipeline as part of the victim system}.
In other words, the adversary cannot simply run an attack algorithm on the model input space.
We thus follow in the direction of \citet{pierazzi_intriguing_2020} and \citet{gao_rethinking_2022} who develop attacks that work end-to-end, as opposed to just attacking a standalone model.
To do this, we develop strategies to ``bypass'' the preprocessors (\cref{sec:attacks}) and to reverse-engineer which preprocessors are being used (\cref{sec:extract}).
Our threat model is:
\begin{itemize}[leftmargin=*,noitemsep,nosep]
    \item The adversary has \emph{black-box, query-based} access to the victim model and can query the model on any input and observe the output label $y \in \mathcal{Y}$.
    The adversary has a limited query budget per input. The adversary knows nothing else about the system. 
    \item The adversary wants to misclassify as many perturbed inputs as possible (either targeted and untargeted), while minimizing the perturbation size---measured by Euclidean distance in \emph{the original input space} $\mathcal{X}_o$.
    \item The victim system accepts inputs of any dimension, and the desired model input size is obtained by cropping and/or resizing as part of an image preprocessing pipeline.
\end{itemize}

\section{Preprocessor-Aware Attacks} \label{sec:attacks}

Decision-based attacks often query a model on many nearby points, e.g., to approximate the local geometry of the boundary.
Since most preprocessors are not \emph{injective} functions, nearby points in the original input space might map onto the same processed image.
Preprocessing thus makes the model's output \emph{invariant} to some input changes.
This can cause the attack to waste queries and prevent it from learning information about the target model.

\begin{figure*}[t]
    \centering
    \includegraphics[width=0.95\textwidth]{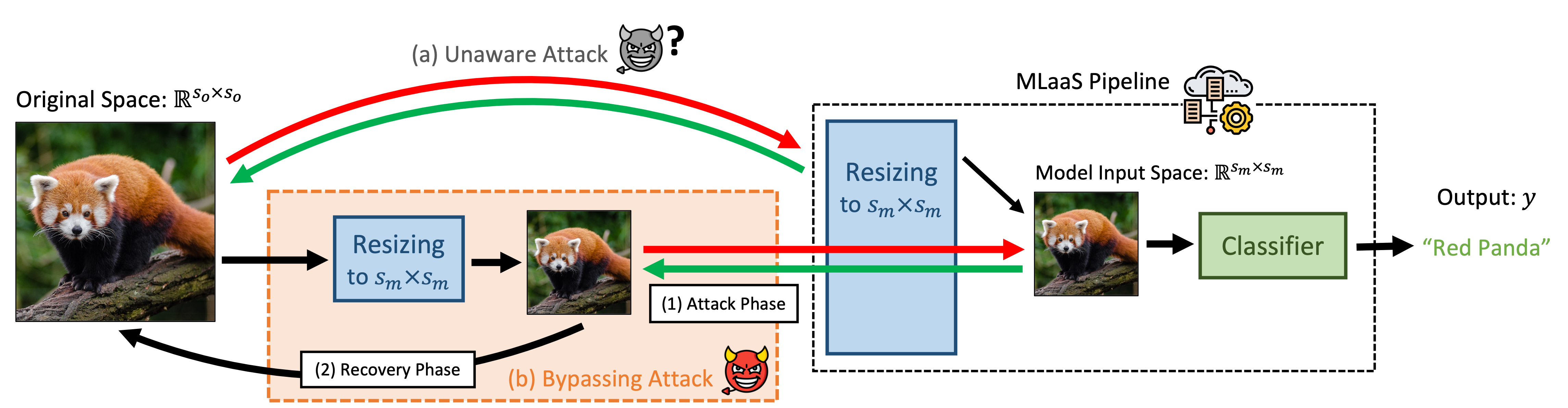}
    \caption{Illustration of our Bypassing Attack with resizing as the preprocessor as a comparison to the unaware or preprocessor-oblivious attack. The red and the green arrows denote the query submitted by the attack and the output returned by the MLaaS pipeline, respectively. The attack phase of our Bypassing Attack first resizes the input image to the correct size used by the target pipeline. This allows any attack algorithm to operate on the model input space directly. The recovery phase then finds the adversarial example in the original space that maps to the one found during the attack phase.}\label{fig:bypass}
\end{figure*}

\subsection{Bypassing Attack} \label{ssec:bypass}

Our \bp Attack in \cref{alg:bypass} avoids these invariances by circumventing the preprocessor entirely.
\cref{fig:bypass} illustrates our attack with a resizing preprocesssor (e.g., $1024 \to 224$).
To allow the \bp Attack to query the model directly, we first map the input image ($x_o \in \mathcal{X}_o$) to the preprocessed space ($t(x_o) \in \mathcal{X}_m$).
Then, in the \emph{Attack Phase}, we execute an off-the-shelf decision-based attack directly on this preprocessed image ($x_m^{\mathrm{adv}} \in \mathcal{X}_m$).

Finally, after completing the attack, we recover the adversarial image in the original space ($x_o^{\mathrm{adv}} \in \mathcal{X}_o$) from $x_m^{\mathrm{adv}}$.
We call this step the \emph{Recovery Phase}.
It finds an adversarial example with minimal perturbation in the original space, by solving the following optimization problem:
\begin{align}
    \argmin_{z_o \in \mathcal{X}_o}~ \norm{z_o - x_o}_2^2 \quad\text{s.t.}\quad t(z_o) = x^{\mathrm{adv}}_m \;. \label{eq:proj}
\end{align}

\begin{algorithm}[t]
    \caption{Outline of Bypassing Attack. This example is built on top of a gradient-approximation-based attack algorithm (e.g., HSJA, QEBA), but it is compatible with any black-box attack. \texttt{ApproxGrad()} and \texttt{AttackUpdate()} are unmodified gradient approximation and perturbation update functions from the base attack. $\mathcal{U}$ is a distribution of vectors on a uniform unit sphere.}
    \label{alg:bypass}
    \begin{algorithmic}
        \STATE {\bfseries Input:} Image $x$, label $y$, classifier $f$, preprocessor $t$
        \STATE {\bfseries Output:} Adversarial examples $x^{\mathrm{adv}}$
        \STATE $x' \leftarrow t(x)$ \COMMENT {Initialization}
        \\
        \COMMENT{\textbf{Attack Phase}: run an attack algorithm of choice}
        \FOR{$i=1$ {\bfseries to} \texttt{num\_steps}}
        \STATE $\tilde X \leftarrow \{x' + \alpha u_b\}_{b=1}^B$ where $u_b \sim \mathcal{U}$
        \STATE $\nabla_{x} S \leftarrow$ \texttt{ApproxGrad}($f \circ t$, $\tilde X$, $y$)
        \STATE $x' \leftarrow$ \texttt{AttackUpdate}($x'$, $\nabla_x S$)
        \ENDFOR
        \\
        \COMMENT{\textbf{Recovery Phase}: exactly recover $x^{\mathrm{adv}}$ in original input space}
        \STATE $x^{\mathrm{adv}} \leftarrow$ \texttt{ExactRecovery}($t, x'$)
        \STATE \textbf{return} $x^{\mathrm{adv}}$
    \end{algorithmic}
\end{algorithm}

\subsubsection{Cropping}

Because almost all image classifiers operate on square images~\citep{wightman_pytorch_2019},
one of the most common preprocessing operations is to first crop the image to a square.
In practice, this means that any pixels on the edge of the image are completely ignored by the classifier.
Our Bypassing Attack exploits this fact by simply removing these cropped pixels, simply running an off-the-shelf attack in the cropped space.
For a more formal statement, see \cref{ap:ssec:bypass_crop}.

\subsubsection{Resizing}

Image resizing is a ubiquitous preprocessing step in any vision system, as
most classifiers are trained only on images of a specific size.
We begin by considering the special case of resizing with ``nearest-neighbor interpolation'', which downsizes images by a factor $k$ simply by selecting only $1$ out of every block of $k$ pixels.
This resize operation is conceptually similar to cropping, and thus the intuition behind our attack is the same:
if we know which pixels are retained by the preprocessor, we can avoid wasting perturbation budget and queries on pixels that are discarded.
Other interpolation methods for resizing, e.g., bilinear or bicubic, work in a similar way, and can all be expressed as a linear transform, i.e., $x_m = t^{\mathrm{res}}(x_o) = M^{\mathrm{res}} x_o$ for $s_o > s_1$.

The attack phase for resizing is exactly the same as that of cropping.
The adversary simply runs an attack algorithm of their choice on the model space $\mathcal{X}_m$.
The main difference comes in the recovery phase which amounts to solving the following optimization problem:
\begin{equation}
    \label{eq:resize_recovery}
    \resizebox{.91\hsize}{!}{%
        $x^{\mathrm{adv}}_o = \argmin_{z_o \in \R^{s_o \times s_o}} \norm{z_o - x_o}_2 ~\text{s.t.}~ M^{\mathrm{res}} z_o = x^{\mathrm{adv}}_m.$%
    }
\end{equation}
\citet{quiring_adversarial_2020,gao_rethinking_2022} solve a similar version of this problem via a gradient-based algorithm.
However, we show that there exists a closed-form solution for the global optimum.
Since the constraint in \eqref{eq:resize_recovery} is an underdetermined linear system, this problem is analogous to finding a minimum-norm solution, given by:
\begin{align}
    x^{\mathrm{adv}}_o = x_o + \delta^*_o = x_o + (M^{\mathrm{res}})^+ \left( x^{\mathrm{adv}}_m - x_m \right).
\end{align}
Here, $(\cdot)^+$ represents the Moore-Penrose pseudo-inverse.
We defer the formal derivation to \cref{ap:ssec:bypass_resize}.

\begin{figure*}[t]
    \centering
    \includegraphics[width=0.95\textwidth]{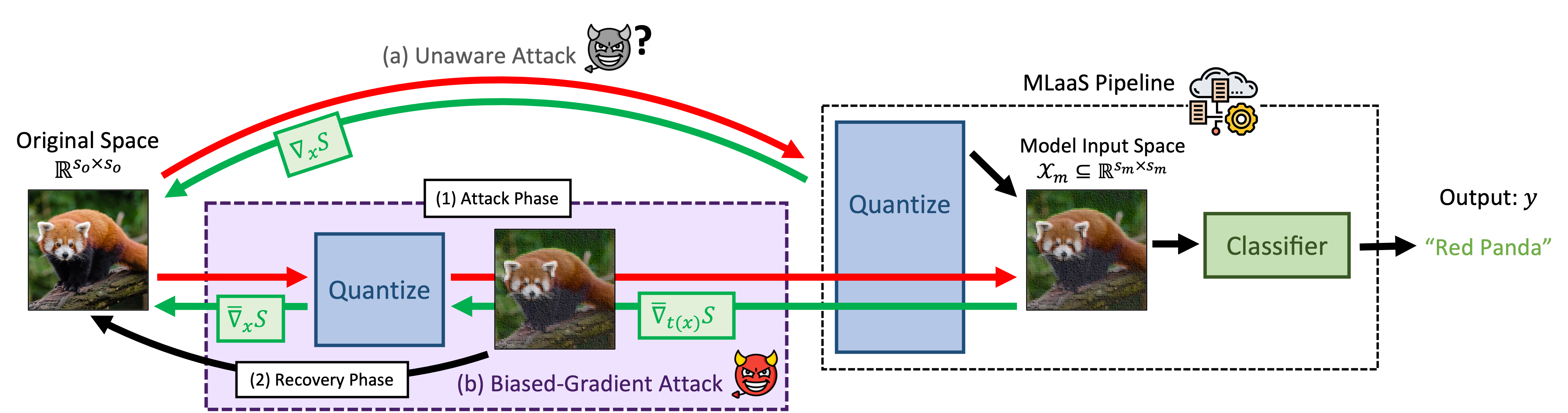}
    \caption{Illustration of the \nbp Attack with quantization as the preprocessor. \nbp Attack cannot directly operate on the model input space like Bypassing Attack. Rather, it takes advantage of the preprocessor knowledge by modifying a specific attack but still operates in the original space.}\label{fig:non_bypass}
\end{figure*}

\textbf{Limitation.}
We have demonstrated how to bypass two very common preprocessors---cropping and resizing---but not all can be bypassed in this way.
Our \bp Attack assumes \textbf{(A1)} the preprocessors are \emph{idempotent}, i.e., $t(t(x)) = t(x)$, and \textbf{(A2)} the preprocessor's output space is continuous.
Most common preprocessing functions are idempotent:
e.g., quantizing an already quantized image makes no difference.
For preprocessors that do not satisfy \textbf{(A2)}, e.g., quantization whose output space is discrete, we propose an alternative attack in the next section.

\subsection{\nbp Attacks} \label{ssec:non_bypass}
We now turn our attention to more general preprocessors that cannot be bypassed without modifying the search space---for example quantization, discretizes a continuous space.
Quantization is one of the most common preprocessors an adversary has to overcome since all common image formats (e.g., PNG or JPEG) discretize the pixel values to 8 bits.
However, prior black-box attacks ignore this fact and operate in the continuous domain.

We thus propose the \nbp Attack in \cref{alg:non_bypass}.
Unlike the \bp Attack, this attack operates in the original space.
Instead of applying a black-box attack as is, the \nbp Attack modifies the base attack in order to bias queries toward directions that the preprocessor is more sensitive to.
The intuition is that while it is hard to completely avoid the invariance of the preprocessor, we can encourage the attack to explore directions that result in large changes in the output space of the preprocessing function.

Our \nbp Attack also consists of an attack and recovery phase.
The attack phase makes two modifications to an underlying gradient approximation attack (e.g., HSJA, QEBA) which we explain below.
The recovery phase simply solves \cref{eq:proj} with a gradient-based method, by relaxing the constraint using a Lagrange multiplier (since closed-formed solutions do not exist in general).
For this, we defer the details to \cref{ap:sec:nbp}.
\cref{fig:non_bypass} illustrates the \nbp Attack for a quantization preprocessor.

\begin{algorithm}[t]
    \caption{Outline of \nbp Attack built on top of gradient-approximation-based attack algorithm.}
    \label{alg:non_bypass}
    \begin{algorithmic}
        \STATE {\bfseries Input:} Image $x$, label $y$, classifier $f$, preprocessor $t$
        \STATE {\bfseries Output:} Adversarial examples $x^{\mathrm{adv}}$
        \STATE $x' \leftarrow x$ \COMMENT {No special initialization}
        \\
        \COMMENT{\textbf{Attack Phase}: run modified attack}
        \FOR{$i=1$ {\bfseries to} \texttt{num\_steps}}
        \STATE\COMMENT{Biased gradient approximation}
        \STATE $\tilde X \leftarrow \{t(x' + \alpha u_b)\}_{b=1}^B$ where $u_b \sim \mathcal{U}$
        \STATE $\nabla_{t(x)} S \leftarrow$ \texttt{ApproxGrad}($f \circ t$, $\tilde X$, $y$)
        \STATE $\Bar{\nabla}_x S \leftarrow \Bar{\nabla}_{t(x)}S \cdot \frac{\partial{t(x)}}{\partial{x}}$ \COMMENT{Backprop through $t$}
        \STATE $x' \leftarrow$ \texttt{AttackUpdate}($x'$, $\nabla_x S$)
        \ENDFOR
        \\
        \COMMENT{\textbf{Recovery Phase}: optimization-based recover $x^{\mathrm{adv}}$ in original space (works for any differentiable $t$)}
        \STATE $x^{\mathrm{adv}} \leftarrow$ \texttt{OptRecovery}($t, x'$)
        \STATE \textbf{return} $x^{\mathrm{adv}}$
    \end{algorithmic}
\end{algorithm}

\noindent \textit{(i) Biased Gradient Approximation:}
We modify the gradient approximation step to account for the preprocessor.
First, consider the adversary's loss function defined as
\begin{equation}
    S(x) \coloneqq \begin{cases}
        \max_{c \in \mathcal{Y} \backslash \{y\}} f_c(x) - f_y(x)     & \text{(untargeted)} \\
        f_{y'}(x) - \max_{c \in \mathcal{Y} \backslash \{y'\}} f_c(x) & \text{(targeted)}
    \end{cases}
\end{equation}
where $(x,y)$ is the input, and $y'\ne y$ is the target label.
Attacks such as HSJA and QEBA estimate the gradient of $S(x)$ by applying finite-differences to the quantity $\phi(x) \coloneqq \text{sign}(S(x))$ which can be measured by querying the model's label.
The attack samples uniformly random unit vectors $\{u_b\}_{b=1}^B$, scales them a hyper-parameter $\alpha$, and computes
\begin{align}
    \nabla_x S(x, \alpha) & \approx \frac{1}{B} \sum_{b=1}^B \phi(t(x + \alpha u_b))u_b \;,
\end{align}
We then perform a change-of-variables to obtain a gradient estimate with respect to $t(x)$ instead of $x$:
\begin{align}
    \frac{1}{B} \sum_{b=1}^B \phi(t(x + \alpha u_b))u_b & = \frac{1}{B} \sum_{b=1}^B \phi(t(x) + \alpha_b' u_b')u_b \label{eq:grad_mod}
\end{align}
where $\alpha_b' = \norm{t(x + \alpha u_b) - t(x)}_2$, and $u_b' = (t(x + \alpha u_b) - t(x)) / \alpha_b'$.
Notice that $\alpha_b' u_b'$ corresponds to a random perturbation in the model space.
Thus, we can ``bypass'' the preprocessor and approximate gradients in the model space instead by substituting $u_b$ with $u_b'$ in \cref{eq:grad_mod}.
\begin{equation}
    \label{eq:grad_of_tx}
    \resizebox{0.91\hsize}{!}{%
        $\Bar{\nabla}_{t(x)} S(x, \alpha) \coloneqq \frac{1}{B} \sum_{b=1}^B \phi(t(x) + \alpha'_b u'_b)u'_b \approx \nabla_{t(x)} S(x, \alpha).$%
    }
\end{equation}
So instead of querying the ML system with inputs $x + \alpha u_b$, we use $t(x + \alpha u_b) = t(x) + \alpha_b' u_b'$ which is equivalent to pre-applying the preprocessor to the queries.
If the preprocessor is idempotent, the model $f$ sees the same processed input in both cases.
This gradient estimator is biased because $u_b'$ depends on $t$.
Concretely, the distribution of $u_b'$ is concentrated around directions that ``survive'' the preprocessor.

\noindent \textit{(ii) Backpropagate Gradients through the Preprocessor:}
The gradient estimate $\Bar{\nabla}_{t(x)} S$ in \eqref{eq:grad_of_tx} is w.r.t. the model space, instead of the original input space where the attack operates.
Hence, we can backpropagate $\Bar{\nabla}_{t(x)} S$ through $t$ according to the chain rule,
$\Bar{\nabla}_x S = \nabla_x t(x) \cdot \Bar{\nabla}_{t(x)} S$
where $\nabla_x t(x)$ is the Jacobian matrix of the preprocessor $t$ w.r.t.~the original space.
In our experiments, we use a differentiable version of quantization and JPEG compression by \citet{shin_jpegresistant_2017} so the Jacobian matrix exists.

\section{Attack Experiments}\label{sec:result}

\subsection{Setup}\label{ssec:exp_setup}

\textbf{Model.}
Similarly to previous works~\citep{brendel_decisionbased_2018}, we evaluate our attacks on a ResNet-18~\citep{he_deep_2016} trained on the ImageNet dataset~\citep{deng_imagenet_2009}.
The model is publicly available in the popular \texttt{timm} package~\citep{wightman_pytorch_2019}.

\textbf{Off-the-shelf attacks.}
We consider four different attacks, Boundary Attack~\citep{brendel_decisionbased_2018}, Sign-OPT~\citep{cheng_signopt_2020}, HopSkipJump Attack (HJSA)~\citep{chen_hopskipjumpattack_2020}, and QEBA~\citep{li_qeba_2020}.
The first three attacks have both targeted and untargeted versions while QEBA is only used as a targeted attack.
We also compare our attacks to the baseline preprocessor-aware attack, SNS~\citep{gao_rethinking_2022}.
As this attack only considers resizing, we adapt it to the other preprocessors we consider.

\textbf{Attack hyperpameters.}
As we discuss in \cref{ssec:hyp} and \cref{ap:ssec:hyp}, a change in preprocessor has a large impact on the optimal choice of hyperparameters for each attack.
We thus sweep hyperparameters for all attacks and report results for the best choice.

\textbf{Metrics.}
We report the average perturbation size ($\ell_2$-norm) of adversarial examples found by each attack---referred to as the ``adversarial  distance'' in short.
Smaller adversarial distance means a stronger attack.

\cref{ap:sec:setup} contains full detail of all our experiments.

\subsection{Bypassing Attack Results}\label{ssec:result}

\textbf{Cropping.}
We consider a common operation that center crops an image of size $256 \times 256$ pixels down to $224 \times 224$ pixels, i.e., $s_o = 256, s_m = 224$.
In \cref{tab:crop_resize}, our \bp approach improves all of the baseline preprocessor-unaware attacks.
The adversarial distance found by the baseline is about 8--16\% higher than that of the \bp Attack counterpart across all settings.
This difference is very close to the portion of the border pixels that are cropped out ($\sqrt{256^2/224^2} - 1 \approx 0.14$), suggesting that the cropping-unaware attacks do waste perturbation on these invariant pixels.
Our \bp Attack also recovers about the same mean adversarial distance as the case where there is no preprocessor (first row of \cref{tab:crop_resize}).

\begin{table*}[t]
    \small
    \caption{Comparing the mean adversarial perturbation norm ($\downarrow$) computed by preprocessor-unaware attacks vs our \bp Attack for the classifier without any preprocessor and with cropping and resizing preprocessors. Lower is better, and the best results for untargeted and targeted settings are in bold.}
    \label{tab:crop_resize}
    \vspace{-5pt}
    \centering
    \begin{tabular}{@{}llrrrrrrr@{}}
        \toprule
        \multirowcell{2}[0ex][l]{Preprocessors} & \multirowcell{2}[0ex][l]{Methods} & \multicolumn{3}{c}{Untargeted Attacks} & \multicolumn{4}{c}{Targeted Attacks}                                                     \\ \cmidrule(lr){3-5} \cmidrule(lr){6-9}                                  &                                        & Boundary                              & Sign-OPT                              & HSJA                                  & Boundary                              & Sign-OPT                              & HSJA          & QEBA \\ \midrule
        None                                    & n/a                               & 4.6                                    & 5.7                                  & \textbf{3.6} & 36.7 & 45.6 & 32.2 & \textbf{19.1} \\ \cmidrule{2-9}
        \multirowcell{2}[0ex][l]{Crop                                                                                                                                                                                   \\($256 \to 224$)} & Unaware & 5.3                                  & 6.5                                  & 4.2                                   & 42.8                                  & 52.7                                  & 38.2                                  & 22.2          \\

                                                & Bypass (ours)                     & 4.6                                    & 5.8                                  & \textbf{3.6} & 37.3 & 46.3 & 32.9 & \textbf{19.6} \\ \cmidrule{2-9}
        \multirowcell{2}[0ex][l]{Resize                                                                                                                                                                                 \\(Nearest)} & Unaware & 21.2                                  & 24.8                                  & 16.5                                  & 172.2                                 & 198.8                                 & 153.4                                 & 90.5          \\
                                                & Bypass (ours)                     & 4.7                                    & 5.8                                  & \textbf{3.7} & 37.7 & 46.3 & 33.3 & \textbf{19.4} \\ \cmidrule{2-9}
        \multirowcell{2}[0ex][l]{Resize                                                                                                                                                                                 \\(Bilinear)} & Unaware & 32.7                                  & 38.2                                  & 25.5                                  & 198.3                                 & 213.0                                 & 188.4                                 & 90.3          \\
                                                & Bypass (ours)                     & 7.4                                    & 9.1                                  & \textbf{6.0} & 58.2 & 70.9 & 50.3 & \textbf{30.0} \\ \cmidrule{2-9}
        \multirowcell{2}[0ex][l]{Resize                                                                                                                                                                                 \\(Bicubic)} & Unaware & 25.7                                  & 29.2                                  & 20.6                                  & 184.8                                 & 207.3                                 & 171.6                                 & 91.2                                  \\
                                                & Bypass (ours)                     & 5.8                                    & 7.1                                  & \textbf{4.5} & 46.4 & 57.7 & 40.6 & \textbf{23.8} \\
        \bottomrule
    \end{tabular}
\end{table*}

\textbf{Resizing.}
We study the three most common interpolation or resampling techniques, i.e., nearest, bilinear, and bicubic.
For an input size of $1024 \times 1024$ (see \cref{tab:crop_resize}), a reasonable image size captured by digital or phone cameras, \textbf{our attack reduces the mean adversarial distance by up to $\bm{4.6\times}$ compared to the preprocessor-oblivious counterpart.}
For all image sizes we experiment with, including 256 and 512 pixels in \cref{tab:resize_all_app}, \bp Attack is always preferable to the resizing-oblivious attack both with and without hyperparameter tuning.

The improvement from the \bp Attack is proportional to the original input dimension.
The benefit diminishes with a smaller original size because the base attack of the Bypassing Attack operates in the model space.
Hence, it minimizes the adversarial distance in that space, i.e., the distance between $x_m^{\mathrm{adv}}$ and $x_m = t(x_o)$.
This distance is likely correlated but not necessarily the same as the true objective distance measured in the original space, i.e., the distance between $x_o^{\mathrm{adv}}$ and $x_o$.
In these cases, it may be preferable to use the \nbp Attack instead.
The results are shown in \cref{tab:bias_grad} and \cref{ssec:non_bypass_result}.

\subsection{\nbp Attack Results} \label{ssec:non_bypass_result}

\begin{table}[t]
    \small
    \caption{Comparison of the mean adversarial perturbation norm ($\downarrow$) found by our \nbp Attacks vs the preprocessor-unaware and the SNS counterparts.}
    \label{tab:bias_grad}
    \vspace{-5pt}
    \centering
    \setlength\tabcolsep{3.0pt}
    \small
    \begin{tabular}{@{}llrrr@{}}
        \toprule
        \multirowcell{2}[0ex][l]{Preprocess} & \multirowcell{2}[0ex][l]{Methods} & \multicolumn{1}{c}{Untg.} & \multicolumn{2}{c}{Targeted}                 \\ \cmidrule(lr){3-3} \cmidrule(lr){4-5}
                                             &                                   & HSJA                      & HSJA                         & QEBA          \\ \midrule
        \multirowcell{3}[0ex][l]{Crop                                                                                                                       \\($256 \to 224$)} & Unaware  & 4.2                                   & 38.2                                  & 22.2                                  \\
                                             & SNS                               & 3.7                       & 35.4                         & 31.5          \\
                                             & Biased-Grad (ours)                & \textbf{3.7}              & 33.1                         & \textbf{19.6} \\ \cmidrule(l){2-5}
        \multirowcell{3}[0ex][l]{Resize                                                                                                                     \\($1024 \to 224$)\\(Nearest)} & Unaware  & 16.5                                   & 153.4                                  & 90.5                                  \\
                                             & SNS                               & 3.9                       & 112.6                        & 32.2          \\
                                             & Biased-Grad (ours)                & \textbf{3.7}              & 23.5                         & \textbf{19.4} \\ \cmidrule(l){2-5}
        \multirowcell{3}[0ex][l]{Quantize                                                                                                                   \\(4 bits)} & Unaware  & 9.7                                   & 63.7                                  & 56.4                                  \\
                                             & SNS                               & 6.4                       & 55.9                         & 57.2          \\
                                             & Biased-Grad (ours)                & \textbf{3.1}              & 39.3                         & \textbf{28.8} \\ \cmidrule(l){2-5}
        \multirowcell{3}[0ex][l]{JPEG                                                                                                                       \\(quality 60)} & Unaware  &       9.2        &  63.2 &  52.7   \\
                                             & SNS                               & 2.7                       & 44.5                         & 44.6          \\
                                             & Biased-Grad (ours)                & \textbf{1.5}              & 25.1                         & \textbf{21.0} \\ \cmidrule(l){2-5}
        \multirowcell{3}[0ex][l]{Neural Compress                                                                                                            \\ \citep{balle_variational_2018} \\ (hyperprior, 8)} & Unaware  &        25.1     &  92.0   &   78.6 \\
                                             & SNS                               & 17.6                      & 83.6                         & 78.9          \\
                                             & Biased-Grad (ours)                & \textbf{15.8}             & \textbf{75.2}                & 75.8          \\ \cmidrule(l){2-5}
        \multirowcell{3}[0ex][l]{Neural Compress                                                                                                            \\ \citep{cheng_learned_2020} \\ (attention, 6)} & Unaware  &        33.8       &  94.1 &   86.9  \\
                                             & SNS                               & 14.3                      & 80.3                         & 75.5          \\
                                             & Biased-Grad (ours)                & \textbf{12.6}             & \textbf{74.8}                & 77.9          \\
        \bottomrule
    \end{tabular}
\end{table}

\begin{table}[t]
    \caption{The mean adversarial distance ($\downarrow$) found by untargeted HSJA on various neural-network-based preprocessors. In the parentheses are the model type and compression level.}\label{tab:neural}
    \vspace{-5pt}
    \small
    \centering
    \setlength\tabcolsep{3.0pt}
    \begin{tabular}{@{}lrrr@{}}
        \toprule
        Neural Preprocessors                           & Unaware & SNS  & Biased-Grad   \\ \midrule
        \citet{balle_variational_2018} (hyperprior, 8) & 25.1    & 17.6 & \textbf{15.8} \\
        \citet{balle_variational_2018} (hyperprior, 6) & 28.7    & 17.0 & \textbf{14.0} \\
        \citet{balle_variational_2018} (factorized, 8) & 24.0    & 15.1 & \textbf{13.9} \\
        \citet{balle_variational_2018} (factorized, 6) & 26.9    & 10.4 & \textbf{11.7} \\
        \citet{cheng_learned_2020} (attention, 6)      & 25.7    & 12.6 & \textbf{14.3} \\
        \citet{cheng_learned_2020} (attention, 4)      & 31.3    & 13.7 & \textbf{13.4} \\
        \citet{cheng_learned_2020} (anchor, 6)         & 27.3    & 8.7  & \textbf{7.0}  \\
        \citet{cheng_learned_2020} (anchor, 4)         & 32.7    & 7.8  & \textbf{6.6}  \\
        SwinIR (denoise level 15)                      & 24.4    & 55.0 & \textbf{10.5} \\ \bottomrule
    \end{tabular}
\end{table}

We evaluate the \nbp Attack on a broad range of preprocessors; in addition to resize and crop, we
include 8/6/4-bit quantization as well as JPEG compression with quality values of 60, 80, and 100.
Moreover, we experiment with neural-network-based compression methods from \citet{balle_variational_2018} and \citet{cheng_learned_2020} as well as SwinIR, a transformer-based denoiser~\citep{liang_swinir_2021}.
We select these methods as representatives of the recent image restoration/compression models which improve upon the traditional computer vision techniques~\citep{zhang_plugandplay_2021,zamir_restormer_2022}.
Importantly, these methods violate our idempotent assumption so they serve an extra purpose of evaluating our attack when the assumption does not hold.

Here, we consider untargeted/targeted HSJA and targeted QEBA as they are consistently the strongest, and the other two do not involve gradient approximation.
From \cref{tab:bias_grad}, \nbp Attack outperforms the preprocessor-unaware counterpart as well as SNS in almost all settings.
A few highlights are:
\textbf{\nbp Attack reduces the mean adversarial distance to only $\bm{\frac{1}{3}}$ and $\bm{\frac{1}{6}}$ of the distance found by the attack without it for 4-bit quantization and JPEG with a quality of 60, respectively.}
The \nbp Attack also outperforms the baselines on neural compression under varying models and compression levels as well as on the SwinIR denoiser (\cref{tab:neural}).
We observe a recurring trend where the benefit of \nbp Attack increases with \emph{stronger} preprocessors, e.g., fewer quantization bits or lower compression quality.

\section{Extracting Preprocessors}\label{sec:extract}

As we have seen,
knowledge of the preprocessor results in much more efficient decision-based attacks.
What is now left is to design a query-efficient attack that actually reverse-engineers the preprocessor used by the target system.

It should not be surprising that this task would be achievable as it is a
particular instance of the more general problem of \emph{model stealing}.
Recent work~\citep{milli2019model,rolnick2020reverse,carlini2020cryptanalytic} has shown a way to completely recover a (functionally-equivalent) neural network using only query access; stealing just a specific part of the model should thus be easier.
Nonetheless, our setting comes with different challenges, both of which relate to the assumed adversary's capabilities:
\begin{enumerate}[leftmargin=*,nosep]
    \item Prior extraction attacks require \emph{high-precision access} to the classifier, i.e., (64-bit) floating-point input/output.
          However, we can only provide valid image files (8-bit) as input and receive only a single decision label as output.
          This invalidates the approaches used in prior work that rely on computing finite differences with epsilon-sized input-output perturbations~\citep{milli2019model}.

    \item Prior attacks needs $10^3$--$10^7$ queries to extract a very simple ($10^3$ parameters) MNIST neural network~\citep{carlini2020cryptanalytic}---in contrast we work with much larger models.
          While the up-front extraction cost can be amortized across many generated adversarial examples, for our attacks to be economically efficient, they must be effective in just a few hundred queries.
\end{enumerate}

\textbf{Intuition.}
Our extraction attack relies on a \emph{guess-and-check strategy}.
Given a hypothesis about the preprocessor (e.g., ``the model uses bilinear
resizing to 224$\times$224 pixels''), we build a set of inputs $Q$ such that the outputs $\{F(q) \mid q \in Q\}$ let us distinguish whether the hypothesis is true or not.
Then, by enumerating a set $\mathbb{T}$ of possible preprocessors, we can use a combination of binary and exhaustive search to reduce this set down to a single preprocessor $t \in \mathbb{T}$.
Our attack (see \cref{alg:extract}) consists of two main components, ``unstable pairs'' and ``pre-images,'' which we describe below.

\subsection{Unstable Example Pairs}

\begin{figure}[t]
    \centering
    \begin{subfigure}[b]{0.21\textwidth}
        \centering
        \includegraphics[width=\textwidth]{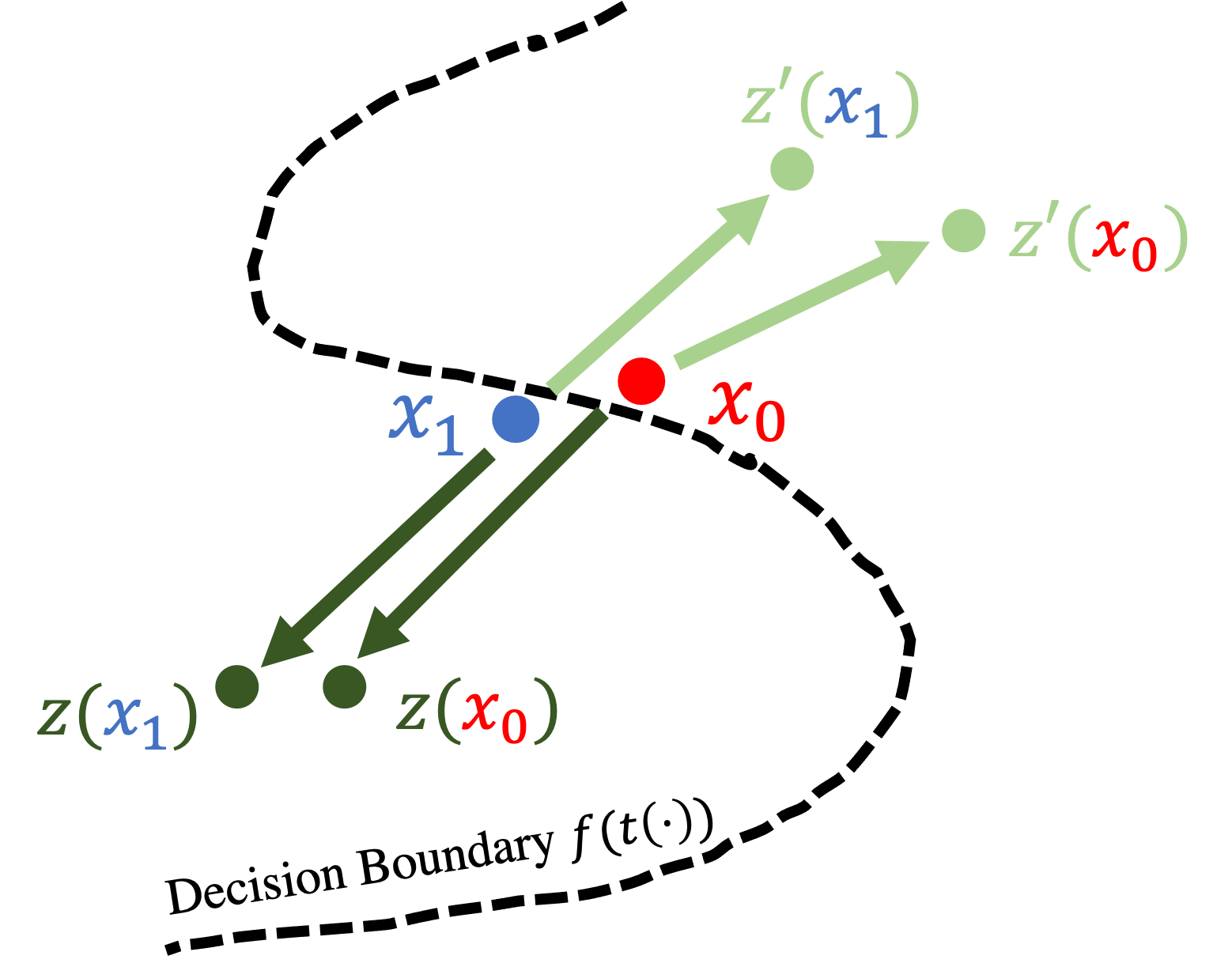}\label{fig:unstable}
    \end{subfigure}
    \hfill
    \begin{subfigure}[b]{0.26\textwidth}
        \centering
        \includegraphics[width=\textwidth]{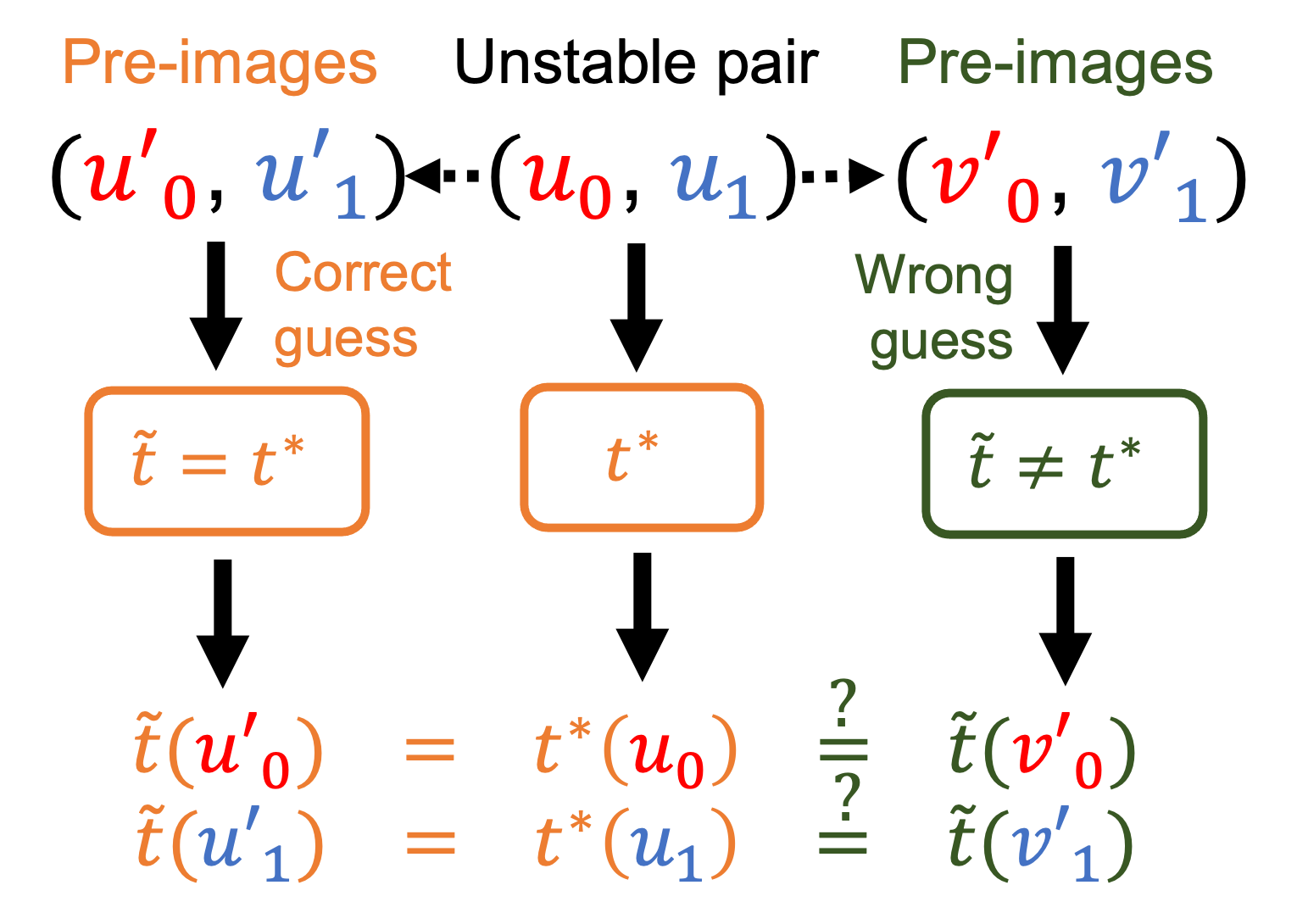}\label{fig:guess}
    \end{subfigure}
    \caption{(Left) An \emph{unstable example pair}, $u_0, u_1$. When slightly perturbed (either $z(\cdot)$ or $z'(\cdot)$), at least one of them is very likely to land on a different side of the decision boundary. (Right) Correct guess on the preprocessor $t^*$ will yield pre-images, $u'_0, u'_1$, that map to the same image as the unstable pair. A wrong guess will not and likely result in a change in at least one of the predictions.}\label{fig:extract}
\end{figure}

The first step of our attack generates an ``unstable pair''.
This is a pair of samples $u_0, u_1$ with two properties: (i) $f(t(u_0)) \ne f(t(u_1))$, and (ii) with high probability $p$, $f(t(z(u_0))) = f(t(z(u_1)))$ for some random perturbation $z : \mathcal{X}_o \to \mathcal{X}_o$.
\cref{fig:unstable} depicts an unstable pair:
the points $x_0 \coloneqq t(u_0)$ and $x_1 \coloneqq t(u_1)$ have opposite labels,
but a small random perturbation $z$ is likely to push both points to the same side of the boundary.
As the perturbation made by $z$ (i.e., $z(u) - u$) grows, $p$ should also increase.

Given two images $x_0$, $x_1$ such that $f(t(x_0)) \ne f(t(x_1))$, we construct an unstable pair $(u_0, u_1)$ by performing two binary searches. %
The first finds a new pair of images $(x'_0, x'_1)$ with $\norm{x'_0 - x'_1}_0 = 1$.
Starting from $(x'_0, x'_1)$, the second binary search finds the unstable pair $(u_0, u_1)$ where $\norm{u_0 - u_1}_\infty = 1$.
This gives us a pair of images that differ by one pixel, and by $1/255$ at that pixel, while also being predicted as different classes.
This process uses only about 40 queries, depending on the input size.
In the interest of space, we provide the detail in \cref{ap:ssec:unstable}.
The rest of the attack uses only the unstable pair; $x_i$ are no longer used. %

\subsection{Hypothesis Testing with Pre-Images} \label{ssec:second_preimage}

Suppose we hypothesize that the preprocessor applied to an
image is some function $\tilde t$ (this is our ``guess'' piece of our guess-and-check attack).
Then, given this unstable example pair $(u_0, u_1)$, we can now implement the ``check'' piece.
For clarity, we denote the actually deployed preprocessor by $t^*$.

We begin by constructing a \emph{pre-image} $u_0' \ne u_0$ so that $\tilde t(u_0) = \tilde t(u_0')$ and analogously for $u_1$ and $u_1'$.
Now if our guess is indeed correct, then it is guaranteed that $\forall i \in \{0,1\}~ F(u'_i) = F(u_i)$ since $\tilde t(u_0') = \tilde t(u_0) = t^*(u_0)$.
On the other hand, if our guess is wrong, then we have $\exists i \in \{0,1\}~ F(u_i') \ne F(u_i)$ with at least some probability $p$.

Let the null hypothesis be $\tilde t \ne t^*$.
When we observe that the predictions of the pre-images do not change, we reject the null hypothesis if $1 - p \le \alpha$ for some threshold $\alpha$ (we choose 0.01).
To increase $p$, we can do two things: (i) simply repeat multiple trials by randomly generating and testing more pre-images and only reject the null hypothesis when \emph{none} of the predictions change, or (ii) increase the size of the perturbation $u_i' - u_i$.
This increases $p$ by definition of the unstable pair.

\subsection{Experiment on Real-World Applications}

\begin{table}[t]
    \small
    \setlength\tabcolsep{4.0pt}
    \centering
    \caption{Number of queries (mean $\pm$ standard deviation) necessary to determine what preprocessor is being used.}
    \label{tab:extract}
    \vspace{-5pt}
    \begin{tabular}{@{}lr@{}}
        \toprule
        Preprocessor Space                          & Num. Queries    \\
        \midrule
        Arbitrary resize (200px--800px)             & $632 \pm 543$   \\
        Arbitrary center crop (0\%-100\%)           & $52.0 \pm 1.3$  \\
        Arbitrary JPEG compression (quality 50-100) & $70.0 \pm 22.8$ \\
        \midrule
        Typical resize (see text)                   & $48.7 \pm 6.8$  \\
        \bottomrule
    \end{tabular}
\end{table}

We use this attack to extract preprocessors for a wide range of models publicly hosted on HuggingFace Hub through their API (\url{https://huggingface.co/docs/api-inference}).
We choose HuggingFace as the preprocessing metadata is available on most models for us to verify our extracted results.
For each experiment, we randomly sample 10 models trained to predict ImageNet-1k classes.
\cref{tab:extract} summarizes the results.

Because our procedure is inherently guess-and-check, we first define the space of all possible preprocessors.
The exact space here depends on the possible knowledge an adversary might have.
In the worst case, an adversary has to enumerate over all possible image sizes ranging from the smallest size used for any image classifier (200px) to the largest size used for any image classifier (800px).
This incurs a cost of 632
queries on average to extract one resizing operator (i.e., both output size and interpolation method).
However, some preprocessors might be more \emph{typical} than others.
We call a preprocessor pipeline ``typical'' if it is used by at least two different models.
For example, ResNet classifiers almost always first resize images to 256 $\times$ 256, and then center-crop the resulting image down to 224 $\times$ 224.
Our set of typical sizes includes 224, 248, 256, 288, 299, 384, and 512 pixels, with bilinear and bicubic interpolations.
With this prior knowledge, the adversary reduces the query cost by over 10 times, down to only $\sim$50 queries

Resize extraction is particularly efficient ($\sim$50 queries) as we can use a binary search to find the crop size, instead of an exhaustive search.
Any wrong guessed crop size larger than the actual size will not change the prediction of the pre-images.
This allows us to run a binary search where the guessed crop size shrinks when the predictions do not change and grows otherwise.
\section{Discussion}\label{sec:discussion}

\subsection{Varying Number of Attack Iterations}\label{ssec:num_steps}

\begin{figure}
    \centering
    \subfloat[Resize (1024$\to$224, nearest)\label{fig:num_steps_resize}]{%
        \includegraphics[width=0.23\textwidth]{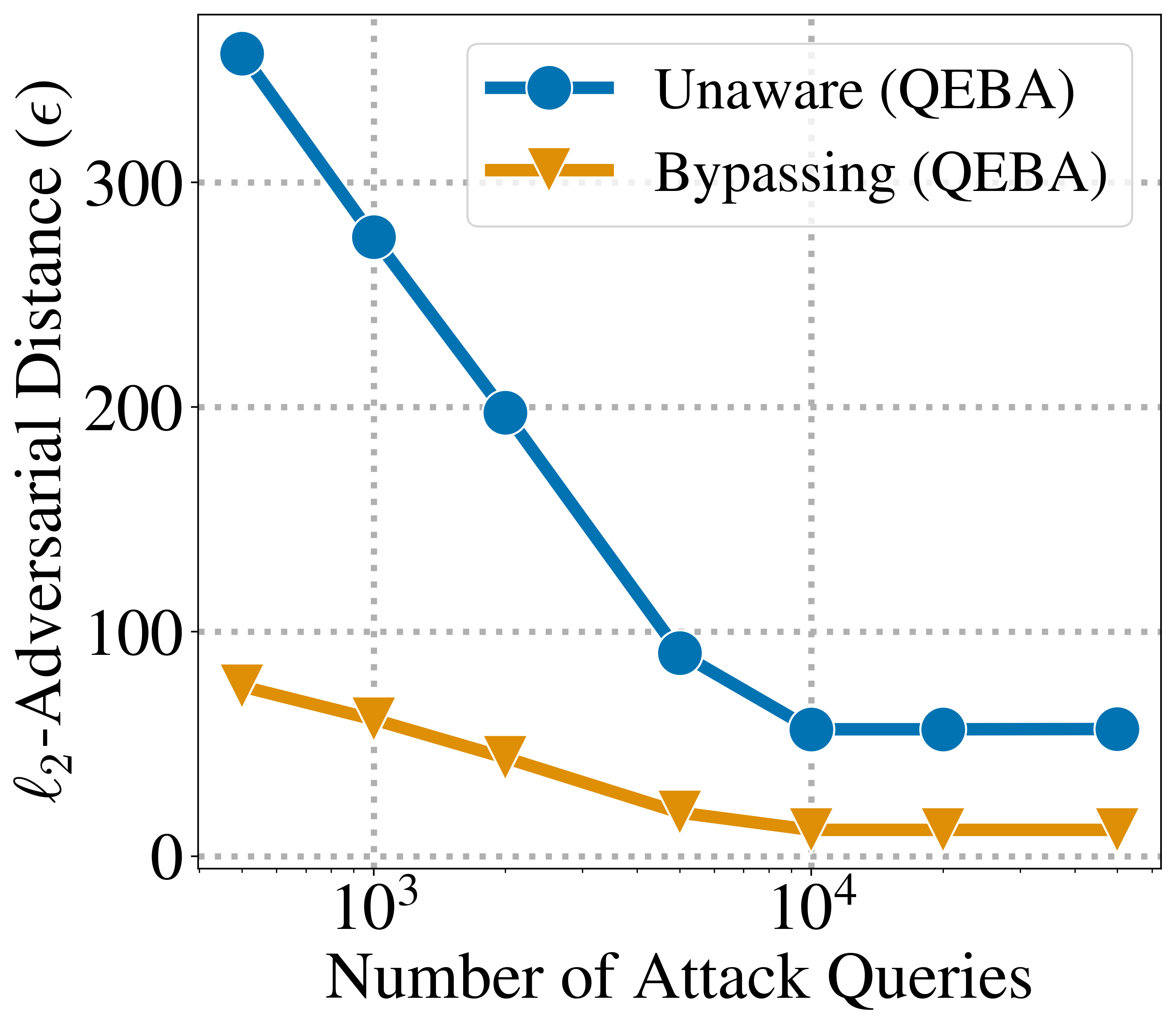}
    }
    \hfill
    \subfloat[JPEG (quality 60)\label{fig:num_steps_jpeg}]{%
        \includegraphics[width=0.23\textwidth]{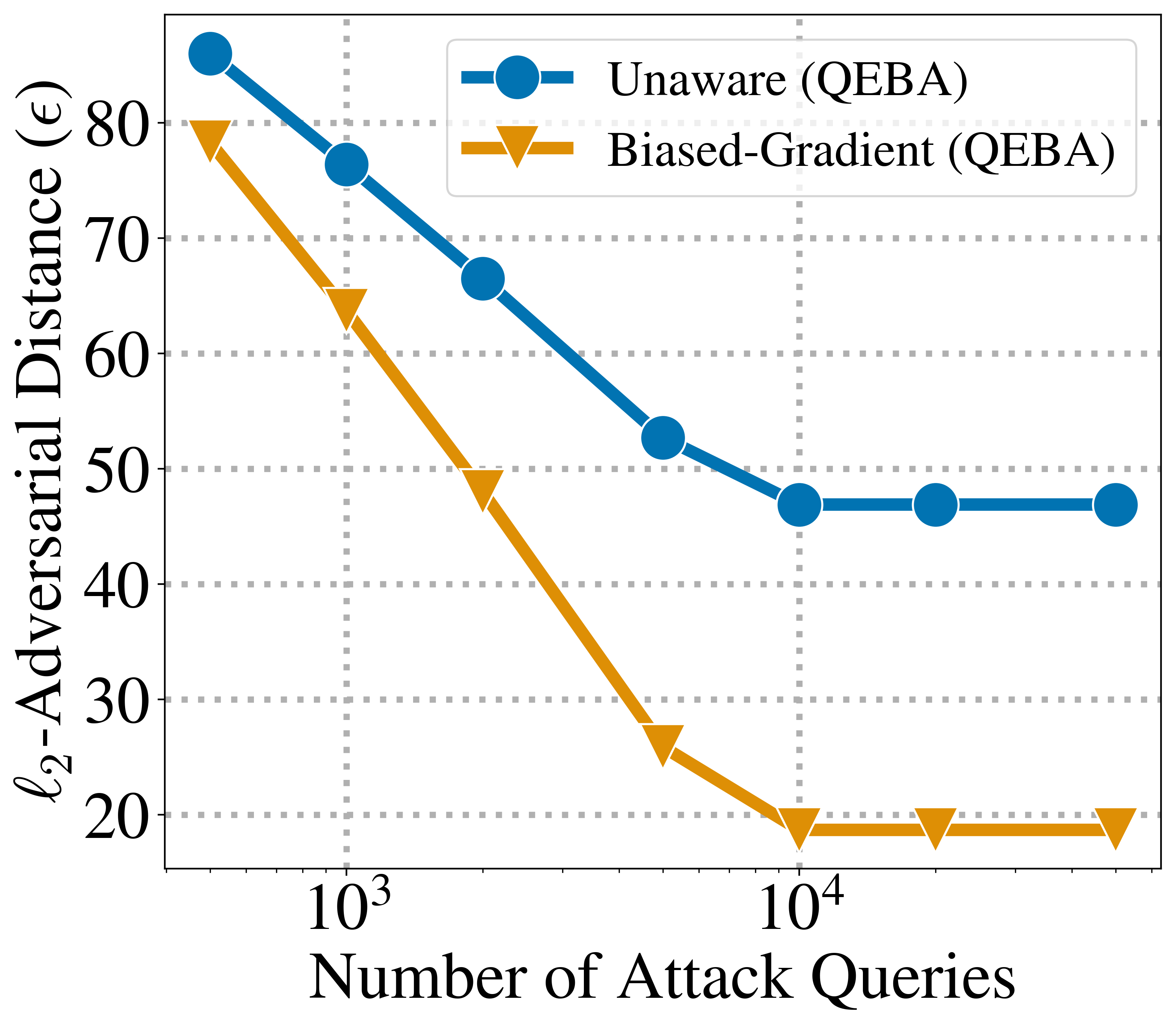}
    }
    \vspace{-5pt}
    \caption{Mean adversarial distance vs the number of queries used by targeted QEBA on two preprocessors.}\label{fig:num_steps}
\end{figure}

\cref{fig:num_steps} plots the mean adversarial distance as a function of the number of queries for QEBA.
Notice that the adversarial distance plateaus after around 10,000 queries, and the distance found by preprocessor-unaware attacks never reaches that of our preprocessor-aware attacks.
This suggests that our attacks do not only improve the efficiency of the algorithms but also allow them to find closer adversarial examples that would have been completely missed otherwise.
See \cref{ap:ssec:num_steps} for more details.

\subsection{Choice of Attack Hyperparameters} \label{ssec:hyp}

Hyperparameter choice is important to the effectiveness of the attacks.
\textbf{In many cases, using the right hyperparameters benefits more than using stronger attack algorithms.}
Choosing the right hyperparameter usually improves the distance found by $\sim$1.5$\times$ and up to 15$\times$ in one case, depending on the attack algorithm.
The knowledge of the preprocessor deployed helps in quickly narrowing down the range of good hyperparameters.
In practice, an adversary would benefit from spending some queries to learn the preprocessors as well as to tune the hyperparameters if one plans to generate many adversarial examples.
For detailed numerical results and discussion, see \cref{ap:ssec:hyp}.

\subsection{Varying the Target Model}\label{ssec:other_models}

We have used the public pre-trained ResNet-18 model as the target model in our experiments, but the conclusion also holds for other models.
We pick two models, EfficientNetV2-Tiny~\citep{tan_efficientnetv2_2021} and DEIT3-Small~\citep{touvron_deit_2022}, with different architectures from ResNet-18, and run the attacks with the resizing (nearest, 1024 $\to$ 288) and JPEG (quality 60) preprocessors, respectively.
The mean adversarial distance for EfficientNetV2-Tiny/DEIT3-Small are 134.2/95.9, 117.7/76.6, \textbf{32.3}/\textbf{48.1} with unaware, SNS, and our \nbp attacks (+ targeted QEBA), respectively.
This corresponds to about 4$\times$ and 2$\times$ improvement over the unaware attack on the two models.

\subsection{Multiple Preprocessors}\label{ssec:multiprep}

\begin{table}[t]
    \caption{Mean adversarial distance ($\downarrow$) found by targeted QEBA when multiple preprocessors are chained together.}\label{tab:multi_prep}
    \vspace{-5pt}
    \small
    \centering
    \setlength\tabcolsep{2.5pt}
    \begin{tabular}{@{}lrrr@{}}
        \toprule
        List of Preprocessors                   & Unaware & SNS   & BG             \\ \midrule
        Resize(1024$\to$256),Crop(224),Quant(8) & 122.4   & 144.9 & \textbf{97.9}  \\
        Resize(1024$\to$256),Crop(224),JPEG(60) & 283.1   & 237.5 & \textbf{153.7} \\
        Resize(512$\to$224),Quant(6)            & 89.8    & 90.9  & \textbf{49.0}  \\
        \bottomrule
    \end{tabular}
\end{table}

\textbf{Preprocessor-aware attacks.}
In practice, multiple preprocessors are used sequentially.
In the case that all the preprocessors can be bypassed, e.g., resizing and cropping, we can bypass the entire pipeline by querying with an appropriate size and padding.
The recovery phase can then be done in the reverse order that the preprocessors are applied.
When at least one preprocessor is not bypassable, we can treat the entire pipeline as one preprocessor and apply the \nbp Attack.
\cref{tab:multi_prep} shows attack results for three common combinations of preprocessors.

\textbf{Extracting multiple preprocessors.}
With the above attack, it becomes trivial to extract multiple preprocessors by extracting each in turn.
Suppose there are two preprocessors $t_1$ and $t_2$, we can first extract $t_1$ by subsuming $t_2$ as part of $f$, i.e., $f' \circ t_1 \coloneqq f \circ t_2 \circ t_1$, and then we move on to guess $t_2$ using the now revealed $t_1$ to construct the pre-images.
In practice, it is actually even easier:
the most common two transformations, resizing and cropping, are almost commutative (i.e., $\texttt{Crop}(\texttt{Resize}(x)) \approx \texttt{Resize}(\texttt{Crop}(x))$ albeit with different crop and resize parameters).
This means that one could either extract cropping or resizing first and still end up with an equivalent overall preprocessor pipeline.
\section{Conclusion}

We have shown that decision-based attacks are sensitive to changes in preprocessors, to a surprising degree.
To develop a strong attack in practice, \textbf{it is more important to get the preprocessor right than to use a stronger attack!}
We propose an extraction attack for commonly used preprocessors and two decision-based attacks aimed to circumvent any preprocessor.
Our approaches are more efficient than the prior work and yield a stronger attack.
We believe that it is important for future work to carefully consider other implicit assumptions in the current adversarial ML literature that may not be true in practice.
We hope that our analysis will inspire future work to further explore this direction.

\subsection*{Acknowledgement}
The authors would like to thank David Wagner for helping with the presentation of the paper, Matthew Jagielski for wonderful discussion on the problem, and Alex Kurakin for comments on early draft of this paper.

A majority of this research was conducted when Chawin was at Google as a student researcher.
For the remaining time at UC Berkeley, Chawin was supported by the Hewlett Foundation through the Center for Long-Term Cybersecurity (CLTC), by the Berkeley Deep Drive project, and by generous gifts from Open Philanthropy.

\bibliography{reference.bib,more_ref.bib}

\begin{thebibliography}{47}
\providecommand{\natexlab}[1]{#1}
\providecommand{\url}[1]{\texttt{#1}}
\expandafter\ifx\csname urlstyle\endcsname\relax
  \providecommand{\doi}[1]{doi: #1}\else
  \providecommand{\doi}{doi: \begingroup \urlstyle{rm}\Url}\fi

\bibitem[Aithal \& Li(2022)Aithal and Li]{aithal2022mitigating}
Aithal, M.~B. and Li, X.
\newblock Mitigating black-box adversarial attacks via output noise perturbation.
\newblock \emph{IEEE Access}, 10:\penalty0 12395--12411, 2022.

\bibitem[Athalye et~al.(2018)Athalye, Carlini, and Wagner]{athalye_obfuscated_2018}
Athalye, A., Carlini, N., and Wagner, D.
\newblock Obfuscated gradients give a false sense of security: Circumventing defenses to adversarial examples.
\newblock In Dy, J. and Krause, A. (eds.), \emph{Proceedings of the 35th International Conference on Machine Learning}, volume~80 of \emph{Proceedings of Machine Learning Research}, pp.\  274--283, {Stockholmsm\"assan, Stockholm Sweden}, July 2018. {PMLR}.

\bibitem[Ball{\'e} et~al.(2018)Ball{\'e}, Minnen, Singh, Hwang, and Johnston]{balle_variational_2018}
Ball{\'e}, J., Minnen, D., Singh, S., Hwang, S.~J., and Johnston, N.
\newblock Variational image compression with a scale hyperprior.
\newblock In \emph{6th International Conference on Learning Representations, {{ICLR}} 2018, Vancouver, {{BC}}, Canada, April 30 - May 3, 2018, Conference Track Proceedings}. {OpenReview.net}, 2018.

\bibitem[B{\'e}gaint et~al.(2020)B{\'e}gaint, Racap{\'e}, Feltman, and Pushparaja]{begaint_compressai_2020}
B{\'e}gaint, J., Racap{\'e}, F., Feltman, S., and Pushparaja, A.
\newblock {{CompressAI}}: A {{PyTorch}} library and evaluation platform for end-to-end compression research.
\newblock \emph{arXiv preprint arXiv:2011.03029}, 2020.

\bibitem[Biggio et~al.(2013)Biggio, Corona, Maiorca, Nelson, {\v S}rndi{\'c}, Laskov, Giacinto, and Roli]{biggio_evasion_2013}
Biggio, B., Corona, I., Maiorca, D., Nelson, B., {\v S}rndi{\'c}, N., Laskov, P., Giacinto, G., and Roli, F.
\newblock Evasion attacks against machine learning at test time.
\newblock In Blockeel, H., Kersting, K., Nijssen, S., and {\v Z}elezn{\'y}, F. (eds.), \emph{Machine {{Learning}} and {{Knowledge Discovery}} in {{Databases}}}, pp.\  387--402, {Berlin, Heidelberg}, 2013. {Springer Berlin Heidelberg}.
\newblock ISBN 978-3-642-40994-3.

\bibitem[Brendel et~al.(2018{\natexlab{a}})Brendel, Rauber, and Bethge]{brendel_decisionbased_2018}
Brendel, W., Rauber, J., and Bethge, M.
\newblock Decision-based adversarial attacks: Reliable attacks against black-box machine learning models.
\newblock In \emph{International Conference on Learning Representations}, 2018{\natexlab{a}}.

\bibitem[Brendel et~al.(2018{\natexlab{b}})Brendel, Rauber, Kurakin, Papernot, Veliqi, Salath{\'e}, Mohanty, and Bethge]{brendel_adversarial_2018}
Brendel, W., Rauber, J., Kurakin, A., Papernot, N., Veliqi, B., Salath{\'e}, M., Mohanty, S.~P., and Bethge, M.
\newblock Adversarial vision challenge.
\newblock Technical report, 2018{\natexlab{b}}.

\bibitem[Carlini \& Wagner(2017)Carlini and Wagner]{carlini_evaluating_2017}
Carlini, N. and Wagner, D.
\newblock Towards evaluating the robustness of neural networks.
\newblock In \emph{2017 {{IEEE}} Symposium on Security and Privacy ({{SP}})}, pp.\  39--57, 2017.
\newblock \doi{10.1109/SP.2017.49}.

\bibitem[Carlini et~al.(2020)Carlini, Jagielski, and Mironov]{carlini2020cryptanalytic}
Carlini, N., Jagielski, M., and Mironov, I.
\newblock Cryptanalytic extraction of neural network models.
\newblock In \emph{Annual International Cryptology Conference}, pp.\  189--218. Springer, 2020.

\bibitem[Chen et~al.(2020)Chen, Jordan, and Wainwright]{chen_hopskipjumpattack_2020}
Chen, J., Jordan, M.~I., and Wainwright, M.~J.
\newblock {{HopSkipJumpAttack}}: A query-efficient decision-based attack.
\newblock \emph{arXiv:1904.02144 [cs, math, stat]}, April 2020.

\bibitem[Chen et~al.(2017)Chen, Zhang, Sharma, Yi, and Hsieh]{chen_zoo_2017}
Chen, P.-Y., Zhang, H., Sharma, Y., Yi, J., and Hsieh, C.-J.
\newblock {{ZOO}}: Zeroth order optimization based black-box attacks to deep neural networks without training substitute models.
\newblock In \emph{Proceedings of the 10th {{ACM}} Workshop on Artificial Intelligence and Security}, {{AISec}} '17, pp.\  15--26, {New York, NY, USA}, 2017. {Association for Computing Machinery}.
\newblock ISBN 978-1-4503-5202-4.
\newblock \doi{10.1145/3128572.3140448}.

\bibitem[Cheng et~al.(2020{\natexlab{a}})Cheng, Singh, Chen, Chen, Liu, and Hsieh]{cheng_signopt_2020}
Cheng, M., Singh, S., Chen, P.~H., Chen, P.-Y., Liu, S., and Hsieh, C.-J.
\newblock Sign-{{OPT}}: A query-efficient hard-label adversarial attack.
\newblock In \emph{International {{Conference}} on {{Learning Representations}}}, 2020{\natexlab{a}}.

\bibitem[Cheng et~al.(2020{\natexlab{b}})Cheng, Sun, Takeuchi, and Katto]{cheng_learned_2020}
Cheng, Z., Sun, H., Takeuchi, M., and Katto, J.
\newblock Learned image compression with discretized gaussian mixture likelihoods and attention modules.
\newblock In \emph{Proceedings of the {{IEEE}} Conference on Computer Vision and Pattern Recognition ({{CVPR}})}, 2020{\natexlab{b}}.

\bibitem[Clarifai()]{clarifai_best_}
Clarifai.
\newblock Best {{NSFW}} model for content detection using {{AI}} | clarifai.
\newblock https://www.clarifai.com/models/nsfw-model-for-content-detection.

\bibitem[Deng et~al.(2009)Deng, Dong, Socher, Li, Li, and {Fei-Fei}]{deng_imagenet_2009}
Deng, J., Dong, W., Socher, R., Li, L.-J., Li, K., and {Fei-Fei}, L.
\newblock {{ImageNet}}: A large-scale hierarchical image database.
\newblock In \emph{2009 {{IEEE}} Conference on Computer Vision and Pattern Recognition}, pp.\  248--255, 2009.
\newblock \doi{10.1109/CVPR.2009.5206848}.

\bibitem[Gao et~al.(2022)Gao, Shumailov, and Fawaz]{gao_rethinking_2022}
Gao, Y., Shumailov, I., and Fawaz, K.
\newblock Rethinking image-scaling attacks: The interplay between vulnerabilities in machine learning systems.
\newblock In Chaudhuri, K., Jegelka, S., Song, L., Szepesvari, C., Niu, G., and Sabato, S. (eds.), \emph{Proceedings of the 39th International Conference on Machine Learning}, volume 162 of \emph{Proceedings of Machine Learning Research}, pp.\  7102--7121. {PMLR}, July 2022.

\bibitem[Goodfellow et~al.(2015)Goodfellow, Shlens, and Szegedy]{goodfellow_explaining_2015}
Goodfellow, I., Shlens, J., and Szegedy, C.
\newblock Explaining and harnessing adversarial examples.
\newblock In \emph{International Conference on Learning Representations}, 2015.

\bibitem[Guo et~al.(2018)Guo, Rana, Cisse, and {van der Maaten}]{guo_countering_2018}
Guo, C., Rana, M., Cisse, M., and {van der Maaten}, L.
\newblock Countering adversarial images using input transformations.
\newblock In \emph{International Conference on Learning Representations}, 2018.

\bibitem[He et~al.(2016)He, Zhang, Ren, and Sun]{he_deep_2016}
He, K., Zhang, X., Ren, S., and Sun, J.
\newblock Deep residual learning for image recognition.
\newblock In \emph{2016 {{IEEE}} Conference on Computer Vision and Pattern Recognition ({{CVPR}})}, pp.\  770--778, 2016.
\newblock \doi{10.1109/CVPR.2016.90}.

\bibitem[Ilyas et~al.(2018)Ilyas, Engstrom, Athalye, and Lin]{ilyas_blackbox_2018}
Ilyas, A., Engstrom, L., Athalye, A., and Lin, J.
\newblock Black-box adversarial attacks with limited queries and information.
\newblock In Dy, J. and Krause, A. (eds.), \emph{Proceedings of the 35th International Conference on Machine Learning}, volume~80 of \emph{Proceedings of Machine Learning Research}, pp.\  2137--2146. {PMLR}, July 2018.

\bibitem[Jagielski et~al.(2020)Jagielski, Carlini, Berthelot, Kurakin, and Papernot]{jagielski_high_2020}
Jagielski, M., Carlini, N., Berthelot, D., Kurakin, A., and Papernot, N.
\newblock High accuracy and high fidelity extraction of neural networks.
\newblock In \emph{29th {{USENIX}} Security Symposium ({{USENIX}} Security 20)}, pp.\  1345--1362. {USENIX Association}, August 2020.
\newblock ISBN 978-1-939133-17-5.

\bibitem[Jha \& Mamidi(2017)Jha and Mamidi]{jha-mamidi-2017-compliment}
Jha, A. and Mamidi, R.
\newblock When does a compliment become sexist? analysis and classification of ambivalent sexism using twitter data.
\newblock In \emph{Proceedings of the Second Workshop on {NLP} and Computational Social Science}, pp.\  7--16, Vancouver, Canada, August 2017. Association for Computational Linguistics.
\newblock \doi{10.18653/v1/W17-2902}.
\newblock URL \url{https://aclanthology.org/W17-2902}.

\bibitem[Li et~al.(2020)Li, Xu, Zhang, Yang, and Li]{li_qeba_2020}
Li, H., Xu, X., Zhang, X., Yang, S., and Li, B.
\newblock {{QEBA}}: Query-efficient boundary-based blackbox attack.
\newblock In \emph{Proceedings of the {{IEEE}}/{{CVF}} Conference on Computer Vision and Pattern Recognition ({{CVPR}})}, June 2020.

\bibitem[Liang et~al.(2021)Liang, Cao, Sun, Zhang, Van~Gool, and Timofte]{liang_swinir_2021}
Liang, J., Cao, J., Sun, G., Zhang, K., Van~Gool, L., and Timofte, R.
\newblock {{SwinIR}}: Image restoration using swin transformer.
\newblock In \emph{Proceedings of the {{IEEE}}/{{CVF}} International Conference on Computer Vision ({{ICCV}}) Workshops}, pp.\  1833--1844, October 2021.

\bibitem[Madry et~al.(2018)Madry, Makelov, Schmidt, Tsipras, and Vladu]{madry_deep_2018}
Madry, A., Makelov, A., Schmidt, L., Tsipras, D., and Vladu, A.
\newblock Towards deep learning models resistant to adversarial attacks.
\newblock In \emph{International Conference on Learning Representations}, 2018.

\bibitem[Milli et~al.(2019)Milli, Schmidt, Dragan, and Hardt]{milli2019model}
Milli, S., Schmidt, L., Dragan, A.~D., and Hardt, M.
\newblock Model reconstruction from model explanations.
\newblock In \emph{Proceedings of the Conference on Fairness, Accountability, and Transparency}, pp.\  1--9, 2019.

\bibitem[{MMEditing Contributors}(2022)]{mmeditingcontributors_mmediting_2022}
{MMEditing Contributors}.
\newblock {{MMEditing}}: {{OpenMMLab}} image and video editing toolbox, 2022.

\bibitem[Papernot et~al.(2017)Papernot, McDaniel, Goodfellow, Jha, Celik, and Swami]{papernot_practical_2017}
Papernot, N., McDaniel, P., Goodfellow, I., Jha, S., Celik, Z.~B., and Swami, A.
\newblock Practical black-box attacks against machine learning.
\newblock In \emph{Proceedings of the 2017 {{ACM}} on Asia Conference on Computer and Communications Security}, {{ASIA CCS}} '17, pp.\  506--519, {New York, NY, USA}, 2017. {Association for Computing Machinery}.
\newblock ISBN 978-1-4503-4944-4.
\newblock \doi{10.1145/3052973.3053009}.

\bibitem[Pierazzi et~al.(2020)Pierazzi, Pendlebury, Cortellazzi, and Cavallaro]{pierazzi_intriguing_2020}
Pierazzi, F., Pendlebury, F., Cortellazzi, J., and Cavallaro, L.
\newblock Intriguing properties of adversarial {{ML}} attacks in the problem space.
\newblock In \emph{2020 {{IEEE Symposium}} on {{Security}} and {{Privacy}} ({{SP}})}, pp.\  1332--1349, May 2020.
\newblock \doi{10.1109/SP40000.2020.00073}.

\bibitem[Qin et~al.(2021)Qin, Fan, Zha, and Wu]{qin2021random}
Qin, Z., Fan, Y., Zha, H., and Wu, B.
\newblock Random noise defense against query-based black-box attacks.
\newblock \emph{Advances in Neural Information Processing Systems}, 34:\penalty0 7650--7663, 2021.

\bibitem[Quiring et~al.(2020)Quiring, Klein, Arp, Johns, and Rieck]{quiring_adversarial_2020}
Quiring, E., Klein, D., Arp, D., Johns, M., and Rieck, K.
\newblock Adversarial preprocessing: Understanding and preventing image-scaling attacks in machine learning.
\newblock In \emph{29th {{USENIX}} Security Symposium ({{USENIX}} Security 20)}, pp.\  1363--1380. {USENIX Association}, August 2020.
\newblock ISBN 978-1-939133-17-5.

\bibitem[Rauber et~al.(2017)Rauber, Brendel, and Bethge]{rauber_foolbox_2017}
Rauber, J., Brendel, W., and Bethge, M.
\newblock Foolbox: A python toolbox to benchmark the robustness of machine learning models.
\newblock \emph{arXiv preprint arXiv:1707.04131}, 2017.

\bibitem[Rolnick \& Kording(2020)Rolnick and Kording]{rolnick2020reverse}
Rolnick, D. and Kording, K.
\newblock Reverse-engineering deep relu networks.
\newblock In \emph{International Conference on Machine Learning}, pp.\  8178--8187. PMLR, 2020.

\bibitem[Shafahi et~al.(2019)Shafahi, Huang, Studer, Feizi, and Goldstein]{shafahi_are_2019}
Shafahi, A., Huang, W.~R., Studer, C., Feizi, S., and Goldstein, T.
\newblock Are adversarial examples inevitable?
\newblock In \emph{International Conference on Learning Representations}, 2019.

\bibitem[Shin \& Song(2017)Shin and Song]{shin_jpegresistant_2017}
Shin, R. and Song, D.
\newblock {{JPEG-resistant}} adversarial images.
\newblock In \emph{Machine {{Learning}} and {{Computer Security Workshop}} (Co-Located with {{NeurIPS}} 2017)}, {Long Beach, CA, USA}, 2017.

\bibitem[Sitawarin et~al.(2022)Sitawarin, {Golan-Strieb}, and Wagner]{sitawarin_demystifying_2022}
Sitawarin, C., {Golan-Strieb}, Z., and Wagner, D.
\newblock Demystifying the adversarial robustness of random transformation defenses.
\newblock In \emph{The {{AAAI-22}} Workshop on Adversarial Machine Learning and Beyond}, 2022.

\bibitem[Song et~al.(2018)Song, Kim, Nowozin, Ermon, and Kushman]{song_pixeldefend_2018}
Song, Y., Kim, T., Nowozin, S., Ermon, S., and Kushman, N.
\newblock {{PixelDefend}}: Leveraging generative models to understand and defend against adversarial examples.
\newblock \emph{arXiv:1710.10766 [cs]}, May 2018.

\bibitem[Szegedy et~al.(2014)Szegedy, Zaremba, Sutskever, Bruna, Erhan, Goodfellow, and Fergus]{szegedy_intriguing_2014}
Szegedy, C., Zaremba, W., Sutskever, I., Bruna, J., Erhan, D., Goodfellow, I., and Fergus, R.
\newblock Intriguing properties of neural networks.
\newblock In \emph{International Conference on Learning Representations}, 2014.

\bibitem[Tan \& Le(2021)Tan and Le]{tan_efficientnetv2_2021}
Tan, M. and Le, Q.
\newblock {{EfficientNetV2}}: Smaller models and faster training.
\newblock In Meila, M. and Zhang, T. (eds.), \emph{Proceedings of the 38th International Conference on Machine Learning}, volume 139 of \emph{Proceedings of Machine Learning Research}, pp.\  10096--10106. {PMLR}, July 2021.

\bibitem[Touvron et~al.(2022)Touvron, Cord, and J{\'e}gou]{touvron_deit_2022}
Touvron, H., Cord, M., and J{\'e}gou, H.
\newblock {{DeiT III}}: Revenge of the {{ViT}}, April 2022.

\bibitem[Tram{\`e}r et~al.(2016)Tram{\`e}r, Zhang, Juels, Reiter, and Ristenpart]{tramer_stealing_2016}
Tram{\`e}r, F., Zhang, F., Juels, A., Reiter, M.~K., and Ristenpart, T.
\newblock Stealing machine learning models via prediction apis.
\newblock In \emph{Proceedings of the 25th {{USENIX}} Conference on Security Symposium}, {{SEC}}'16, pp.\  601--618, {USA}, 2016. {USENIX Association}.
\newblock ISBN 978-1-931971-32-4.

\bibitem[Tram{\`e}r et~al.(2019)Tram{\`e}r, Dupr{\'e}, Rusak, Pellegrino, and Boneh]{tramer_adversarial_2019a}
Tram{\`e}r, F., Dupr{\'e}, P., Rusak, G., Pellegrino, G., and Boneh, D.
\newblock {{AdVersarial}}: Perceptual ad blocking meets adversarial machine learning.
\newblock In \emph{Proceedings of the 2019 {{ACM SIGSAC Conference}} on {{Computer}} and {{Communications Security}}}, pp.\  2005--2021, November 2019.
\newblock \doi{10.1145/3319535.3354222}.

\bibitem[Tramer et~al.(2020)Tramer, Carlini, Brendel, and Madry]{tramer_adaptive_2020}
Tramer, F., Carlini, N., Brendel, W., and Madry, A.
\newblock On adaptive attacks to adversarial example defenses.
\newblock In Larochelle, H., Ranzato, M., Hadsell, R., Balcan, M.~F., and Lin, H. (eds.), \emph{Advances in Neural Information Processing Systems}, volume~33, pp.\  1633--1645. {Curran Associates, Inc.}, 2020.

\bibitem[Waseem et~al.(2017)Waseem, Davidson, Warmsley, and Weber]{waseem-etal-2017-understanding}
Waseem, Z., Davidson, T., Warmsley, D., and Weber, I.
\newblock Understanding abuse: A typology of abusive language detection subtasks.
\newblock In \emph{Proceedings of the First Workshop on Abusive Language Online}, pp.\  78--84, Vancouver, BC, Canada, August 2017. Association for Computational Linguistics.
\newblock \doi{10.18653/v1/W17-3012}.
\newblock URL \url{https://aclanthology.org/W17-3012}.

\bibitem[Wightman(2019)]{wightman_pytorch_2019}
Wightman, R.
\newblock {{PyTorch}} image models.
\newblock GitHub, 2019.

\bibitem[Zamir et~al.(2022)Zamir, Arora, Khan, Hayat, Khan, and Yang]{zamir_restormer_2022}
Zamir, S.~W., Arora, A., Khan, S., Hayat, M., Khan, F.~S., and Yang, M.-H.
\newblock Restormer: Efficient transformer for high-resolution image restoration.
\newblock In \emph{Proceedings of the {{IEEE}}/{{CVF}} Conference on Computer Vision and Pattern Recognition ({{CVPR}})}, pp.\  5728--5739, June 2022.

\bibitem[Zhang et~al.(2021)Zhang, Li, Zuo, Zhang, Van~Gool, and Timofte]{zhang_plugandplay_2021}
Zhang, K., Li, Y., Zuo, W., Zhang, L., Van~Gool, L., and Timofte, R.
\newblock Plug-and-play image restoration with deep denoiser prior.
\newblock \emph{IEEE Transactions on Pattern Analysis and Machine Intelligence}, 44\penalty0 (10):\penalty0 6360--6376, 2021.

\end{thebibliography}
\bibliographystyle{icml2023}

\newpage
\appendix
\onecolumn
\section{Detailed Experiment Setup} \label{ap:sec:setup}

We use a pre-trained ResNet-18 model from a well-known repository \texttt{timm}~\citep{wightman_pytorch_2019} which is implemented in PyTorch and trained on inputs of size $224 \times 224$.
This model is fixed throughout all the experiments.
The experiments are run on multiple remote servers with either Nvidia Tesla A100 40GB or Nvidia V100 GPUs.

Implementations of Boundary Attack and HSJA are taken from the \texttt{Foolbox} package~\citep{rauber_foolbox_2017}.\footnote{We use code from the commit: \url{https://github.com/bethgelab/foolbox/commit/de48acaaf46c9d5d4ea85360cadb5ab522de53bc}.}
For Sign-OPT Attack and QEBA, we use the official, publicly available implementation.\footnote{Sign-OPT attack: \url{https://github.com/cmhcbb/attackbox}. QEBA: \url{https://github.com/AI-secure/QEBA}.}

To compare effectiveness of the attacks, we report the average perturbation size ($\ell_2$-norm) of the adversarial examples computed on 1,000 random test samples.
We will refer to this quantity as the adversarial distance in short.
Smaller adversarial distance means a stronger attack.
Unless stated otherwise, all the attacks use 5,000 queries per one test sample.

The implementation of the neural compression models along with their weights is taken from \citet{begaint_compressai_2020}.
These models only accept input sizes that are powers of two so we change the input size from the default 224$\times$224 to 256$\times$256 here.
For SwinIR, we use the publicly available code and weights from \citet{mmeditingcontributors_mmediting_2022}.
Backpropagating the gradients through a SwinIR model takes up a large amount of memory which we cannot fit into our GPUs.
So we have to reduce the input size to 128$\times$128 pixels.
Note that due to this mismatch in the original input sizes, it is not recommended to compare the mean adversarial distances across these preprocessors.

\subsection{Hyperparameter Sweep} \label{ap:ssec:setup_hyp}

We find that the choice of hyperparameters of the four attack algorithms plays an important role in their effectiveness, and it is not clear how an attacker would know apriori how to choose such hyperparameters.
In reality, the adversary would benefit from spending some queries to tune the hyperparameters on a few samples.
Coming up with the most efficient tuning algorithm is outside of the scope of this work.
Nonetheless, we account for this effect by repeating all experiments with multiple choices of hyperparameters and reporting the results with both the best sets throughout the paper.
We include some of the results with both the best and the default hyperparameters for comparisons in \cref{tab:crop_app}, \cref{tab:quantize}, and \cref{tab:jpeg}.

For Boundary attack, we sweep the two choices of step size, one along the direction towards the original input and the other in the orthogonal direction.
The default values are $(0.01, 0.01)$, respectively, and the swept values are $(0.1, 0.01)$, $(0.001, 0.01)$, $(0.01, 0.1)$, and $(0.01, 0.001)$.

For Sign-OPT attack, we consider the update step size $\alpha$ and the gradient estimate step size $\beta$.
Their default values are $(0.2, 0.001)$ respectively, and we sweep the following values: $(0.2, 0.01)$, $(0.2, 0.0001)$, $(0.02, 0.001)$, and $(2, 0.01)$.

We only tune one hyperparameter for HSJA and QEBA attacks but with the same number of settings (five) as the other two attacks above.
For HSJA, we tune the update step size $\gamma$ by trying values of $10^1$ (default), $10^2$, $10^3$, $10^4$, and $10^5$.
The optimal value of $\gamma$ is always at a higher range than $10^1$, not smaller.
Lastly, we search the ratio $r$ that controls the latent dimension that QEBA samples its random noise from for gradient approximation.
We search over $r=2,4,8,16,32$.

The observed trends and the recommended hyperparameters are discussed further below in \cref{ap:ssec:hyp}.
\section{Bypassing Attacks}

Here, we provide additional details on the \bp Attack for cropping and resizing preprocessors.

\subsection{Cropping Preprocessor} \label{ap:ssec:bypass_crop}

\paragraph{Attack Phase for Cropping.}
To bypass the cropping transformation, the attacker simply submits an already cropped input and runs any query-based attack algorithm in the space $\R^{s_m \times s_m}$ instead of $\R^{s_o \times s_o}$.
Without any modification to the attack algorithm, it is able to operate directly on the model space as if there is no preprocessing.

\paragraph{Recovery Phase for Cropping.}
In order for the adversarial example obtained from the attack phase to be useful in input space, the adversary still has to produce an adversarial example in the original space with the smallest possible Euclidean distance to the original input.
It should be obvious that for cropping, this operation simply equates to padding this adversarial example with the original edge pixels.

\paragraph{Formal Definition of Cropping's Recovery Phase.}
We now formally describe what it means to crop an image.
Given an input image of size $s_o \times s_o$, a crop operation removes the edge pixels of any image larger than a specified size, denoted by $s_m \times s_m$, such that the output has the size $s_o \times s_o$.
Given an (flattened) input image $x_o \in \R^{s_o \times s_o}$ and the cropped image $x_m \in \R^{s_m \times s_m}$, we can write cropping as the following linear transformation, when $s_o > s_m$,
\begin{align}
  x_m & = M^{\mathrm{crop}} x_o
\end{align}
where $M^{\mathrm{crop}} \in \R^{s_o^2 \times s_m^2}$ is a sparse binary matrix.
Each row of $M^{\mathrm{crop}}$ has exactly one entry being $1$ at a position of the corresponding non-edge pixel while the rest are $0$.
Note that we drop the ``color-channel'' dimension for simplicity since most of the preprocessors in this paper is applied channel-wise.
We are only interested in the scenario when $s_o > s_m$ because otherwise, the preprocessing simply becomes an identity function.

Let the adversarial example in the model space as obtained from the attack phase be $x_m^{\mathrm{adv}} \in \R^{s_m \times s_m}$.
The adversary can recover the corresponding adversarial example in the original space, $x_o^{\mathrm{adv}} \in \R^{s_o \times s_o}$, by padding $x_m^{\mathrm{adv}}$ with the edge pixels of $x_o$.

It is simple to show that $x_o^{\mathrm{adv}}$ is a projection of $x_o$ onto the set $\mathcal{T}^{\mathrm{crop}}(x_m^{\mathrm{adv}}) \coloneqq \{x \in \mathcal{X}_o \mid t^{\mathrm{crop}}(x) = x_m^{\mathrm{adv}}\}$, i.e.,
\begin{align}
  x_o^{\mathrm{adv}} = \argmin_{x \in \mathcal{T}^{\mathrm{crop}}(x_m^{\mathrm{adv}})}~\norm{x - x_o}_2^2 \label{eq:opt_crop}
\end{align}
\begin{proof}
  We can split $\norm{x - x_o}_2^2$ into two terms
  \begin{align}
    \sum_{i \in E}\left( [x]_i - [x_o]_i \right)^2 + \sum_{i \notin E}\left( [x]_i - [x_o]_i \right)^2 \label{eq:crop_proof}
  \end{align}
  where $E$ is a set of edge pixel indices.
  The second term is fixed to $\norm{x_m^{\mathrm{adv}} - t^{\mathrm{crop}}(x_o)}_2^2$ for any $x \in \mathcal{T}^{\mathrm{crop}}(x_m^{\mathrm{adv}})$.
  When $x = x_o^{\mathrm{adv}}$, the first term is zero because $x_o^{\mathrm{adv}}$ is obtained by padding $x_m^{\mathrm{adv}}$ with the edge pixels of $x_o$.
  Since the first term is non-negative, we know that $x_o^{\mathrm{adv}}$ is a unique global minimum of \eqref{eq:opt_crop}.
\end{proof}

\subsection{Resizing Preprocessor} \label{ap:ssec:bypass_resize}

\subsubsection{Computing the Transformation Matrix $M^{\mathrm{res}}$}

Not all image resizing operations are the same;
the main step that varies between them is called the ``interpolation'' mode.
Interpolation determines how the new pixels in the resized image depend on (multiple) pixels in the original image.
Generally, resizing represents some form of a weighted average.
How the weights are computed and how many of the original pixels should be used varies by specific interpolation methods.

For nearest interpolation (zeroth order), $M^{\mathrm{res}}$ is a sparse binary matrix with exactly one $1$ per row.
For higher-order interpolations, a pixel in $x_m$ can be regarded as a weighted average of certain pixels in $x_o$.
Here, $M^{\mathrm{res}}$ is no longer binary, and each of its rows represents these weights which are between $0$ and $1$.
For instance, since one pixel in a bilinear resized image is a weighted average of four pixels ($2 \times 2$ pixels) in the original image, $M^{\mathrm{res}}$ for bilinear interpolation has four non-zero elements per row.
On the other hand, $M^{\mathrm{res}}$ for bicubic interpolation has 16 non-zero elements per row ($4 \times 4$ pixels).
$M^{\mathrm{res}}$ is still generally sparse for $s_o > s_1$ and is more sparse when $s_o/s_1$ increases.

The matrix $M^{\mathrm{res}}$ can be computed analytically for any given $s_o$ and $s_1$.
Alternatively, it can be populated programmatically, by setting each pixel in the original image to 1, one at a time, then performing the resize, and gathering the output.
This method is computationally more expensive but simple, applicable to any sampling order, and robust to minor differences in different resizing implementations.

\subsubsection{Recovery Phase for Resizing}

The recovery phase involves some amount of linear algebra, as it is equivalent to solving the following linear system of equations
\begin{align}
  x^{\mathrm{adv}}_m = M^{\mathrm{res}} x^{\mathrm{adv}}_o. \label{eq:res_eq}
\end{align}
to find $x^{\mathrm{adv}}_o$.
Note that for $s_o > s_m$, this is an underdetermined system so there exist multiple solutions.
A minimum-norm solution, $x^*_o$, can be obtained by computing the right pseudo-inverse of $M^{\mathrm{res}}$ given by
\begin{align}
  (M^{\mathrm{res}})^+ & = (M^{\mathrm{res}})^\top (M^{\mathrm{res}} (M^{\mathrm{res}})^\top)^+ \label{eq:res_pinv} \\
  x^*_o                & = (M^{\mathrm{res}})^+ x^{\mathrm{adv}}_m \label{eq:res_sol}
\end{align}

However, the adversary does not want to find a minimum-norm original sample $x^*_o$ but rather a minimum-norm perturbation $\delta^*_o = x^{\mathrm{adv}}_o - x_o$.
This can be accomplished by modifying \eqref{eq:res_eq} and \eqref{eq:res_sol} slightly
\begin{align}
  M^{\mathrm{res}} \left( x_o + \delta^*_o \right) & = x^{\mathrm{adv}}_m                                                                    \\
  M^{\mathrm{res}} \delta^*_o                      & = x^{\mathrm{adv}}_m - M^{\mathrm{res}} x_o                                             \\
  \delta^*_o                                       & = (M^{\mathrm{res}})^+ \left( x^{\mathrm{adv}}_m - M^{\mathrm{res}} x_o \right)         \\
  \delta^*_o                                       & = (M^{\mathrm{res}})^+ \left( x^{\mathrm{adv}}_m - x_m \right).  \label{eq:res_recover}
\end{align}

\eqref{eq:res_recover} summarizes the recovery phase for resizing.
By construction, it guarantees that $\delta^*_o$ is a minimum-norm perturbation for a given $x^{\mathrm{adv}}_m$, or $x^{\mathrm{adv}}_o = x_o + \delta^*_o$ is a projection of $x_o$ onto the set of solutions that map to $x^{\mathrm{adv}}_m$ after resizing.
In other words, by replacing any $\delta_o$ with $z_o - x_o$, we have
\begin{align}
  x^{\mathrm{adv}}_o                    & = \argmin_{z_o \in \R^{s_o \times s_o}}~ \norm{z_o - x_o}_2 \\
  \text{s.t.}\quad M^{\mathrm{res}} z_o & = x^{\mathrm{adv}}_m.
\end{align}

In practice, we can compute $\delta_o^*$ by either using an iterative solver on \eqref{eq:res_eq} directly, or by pre-computing the pseudo-inverse in \eqref{eq:res_pinv}.
The former does not require caching any matrix but must be recomputed for every input.
Caching the pseudo-inverse is more computationally expensive but is done only once.
Since $M^{\mathrm{res}}$ is sparse, both options are very efficient.
\section{\nbp Attack}\label{ap:sec:nbp}

\subsection{Details on the Recovery Phase}

We propose a recovery phase for general preprocessors which should also work for cropping and resizing as well, albeit less efficiently compared to the one in Bypassing Attack.
Assuming that the preprocessor is differentiable or has a differentiable approximation, it is possible to replace the exact projection mechanism for finding $x^{\mathrm{adv}}_o$ with an iterative method.
Specifically, consider relaxing the constraint from \eqref{eq:proj} with a Lagrange multiplier:
\begin{align}
    \argmin_{z_o \in \mathcal{X}_o}~ \norm{z_o - x_o}_2^2 + \lambda \norm{t(z_o) - x^{\mathrm{adv}}_m}_2^2.
\end{align}

This optimization problem can then be solved with gradient descent combined with a binary search on the Lagrange multiplier $\lambda$.
We emphasize that, unlike the exact recovery for resizing or cropping, the second term does not necessarily need to be driven down to zero, i.e., $t(z^*_o)=x^{\mathrm{adv}}_m$.
For the \nbp Attack, $x^{\mathrm{adv}}_m$ can be seen as a proxy to make $z^*_o$ misclassified by $f(t(\cdot))$ or as a guide to move $t(z_o)$ towards.
Specifically, we want the smallest $\lambda$ such that the solution $z^*_o$ minimizes $\norm{z^*_o - x_o}_2$ while also being misclassified.

To this end, we use binary search on $\lambda$ by increasing/decreasing it when $z^*_o$ is correctly/incorrectly classified.
Throughout this paper, we use 10 binary search steps (3 steps in the case of the neural-network-based preprocessor as computing gradients through these models can be expensive).
Each step only requires exactly one query to check the predicted label at the end.
In practice, we also impose a constraint that keeps $z_0$ in the input domain $[0,1]$ using a change of variable trick inspired by the attack from \citet{carlini_evaluating_2017}.
\section{Preprocessor Extraction Attacks}\label{ap:sec:extract}

\begin{algorithm}[t]
    \caption{Outline of our extraction attack. Here, we denote the target classification pipeline as $F \coloneqq f \circ t$ where $f$ and $t$ are unknown to the adversary.}\label{alg:extract}
    \begin{algorithmic}
        \STATE {\bfseries Input:} Target classifier API $F$, a pair of images $x_0, x_1$ where $F(x_0) \ne F(x_1)$, set of guessed preprocessors $\mathbb{T}$.
        \STATE {\bfseries Output:} Extracted preprocessor $t^*$
        \STATE $(u_0, u_1) \leftarrow$ \texttt{GenUnstablePair}($F$, $(x_0, x_1)$)
        \\
        \COMMENT{(Optional) Estimate $p$ to compute \texttt{num\_trials} given the desired $p$-value $\alpha$} \\
        \texttt{num\_trials} $\leftarrow$ \texttt{GetNumTrials}($F$, $Q$, $p$, $\alpha$)
        \FOR{$t \in \mathbb{T}$}
        \FOR{$i=1$ {\bfseries to} \texttt{num\_trials}}
        \STATE $(u'_0, u'_1) \leftarrow$ \texttt{GenPreImage}($t, (u_0, u_1)$)
        \STATE $c_i \leftarrow F(u'_0) = F(u_0) ~\&~ F(u'_1) = F(u_1)$
        \ENDFOR
        \IF{$\forall_i c_i = 1$}
        \STATE \textbf{return} $t$
        \ENDIF
        \ENDFOR
    \end{algorithmic}
\end{algorithm}

\subsection{Detailed Construction of Unstable Pairs}\label{ap:ssec:unstable}
We begin by identifying (any) two images $x_0$, $x_1$ such that $f(t(x_0)) \ne f(t(x_1))$.
This step should be easy: it suffices to identify two valid images that actually belong to different classes, or to make random (large-magnitude) modifications to one image $x_0$ until it switches classes and then call the perturbed image $x_1$.
Intuitively, because $f(t(x_0)) \ne f(t(x_1))$, if we were to interpolate between
$x_0$ and $x_1$, there must be a midpoint $\bar x$ where the decision changes.
By picking $x_0$ and $x_1$ to straddle this midpoint $\bar x$, we obtain an unstable example pair.
If the input space of the pipeline were continuous, we can generate an unstable pair, up to the floating-point precision, with a single binary search.
However, since we focus on real systems that accept only 8-bit images, we need to take multiple extra steps to create the pair that differs by only one bit on one pixel.

First, we reduce the $\ell_0$ difference between the two images via binary search.
Construct a new image $\bar x$ where each pixel is independently chosen (uniformly at random) as the pixel value either from the image $x_0$ or from the image $x_1$.
This new image $\bar x$ now roughly shares half of the pixels with $x_0$ and half of the pixels with $x_1$.
If $f(t(\bar x)) = f(t(x_0))$ replace $x_0$ with $\bar x$ and repeat;
if $f(t(\bar x)) = f(t(x_1))$ then replace $x_1$ with $\bar x$ and repeat.

Next, reduce the $\ell_\infty$ difference between these two images, again following the same binary search procedure.
Let $\bar x = (x_0 + x_1)/2$, and query the model to obtain $f(t(\bar x))$.
If $f(t(\bar x)) = f(t(x_0))$ then replace $x_0$ with $\bar x$ and repeat;
if $f(t(\bar x)) = f(t(x_1))$ then replace $x_1$ with $\bar x$ and repeat.
Do this until $x_0$ and $x_1$ differ from each other by at most 1/255 (the smallest difference two images can have).
This will eventually give a pair of images that now differ in exactly one pixel coordinate, and in this one coordinate by exactly 1/255.
By construction, these two images are also classified as different classes by the pipeline $F$.
We call them an unstable pair.
Note that we have not relied on the knowledge of $t$ as we have only treated $f \circ t$ as a single function.

\subsection{Detailed Pre-Image Attack} \label{ap:ssec:second_preimage}

Once we obtain the unstable pair $(u_0, u_1)$, the next step is to use them to generate multiple pre-image $(u'_0, u'_1)$ and ``check'' our ``guess'' of the preprocessor.
Before explaining how the pre-images are generated, we will first expand on the implications of the two outcomes: our guess is either right or wrong.

\textit{Our guess is correct:}
In the case that our guess is right, ($\tilde t = t^*$), the following equality will hold for $i \in \{0,1\}$,
\begin{equation} \label{eqn:sim}
    f(t^*(u_i')) = f(\tilde t(u_i')) = f(\tilde t(u_i)) = f(t^*(u_i))
\end{equation}
where the first equality holds by the assumption that $\tilde t = t^*$, the second equality holds by construction that $u_i'$ and $u_i$ are pre-images, and the final equality holds under the first correctness assumption.
From here, we can conclude
\begin{equation*}
    \overunderbraces{&&\br{3}{\text{By construction}}}%
    {&f(t^*(u_0'))=&f(t^*(u_0))&\ne&f(t^*(u_1))&=f(t^*(u_1'))}%
    {&\br{2}{\text{By \eqref{eqn:sim}}}& &\br{2}{\text{By \eqref{eqn:sim}}}}.
\end{equation*}
Put simply, this means that if we feed the pipeline with $u_0'$ and $u_1'$, and if our preprocessor guess is correct, then the pipeline will give two different answers $f(t^*(u_0')) \ne f(t^*(u_1'))$.

\textit{Our guess is wrong:}
On the other hand, if our guess at the preprocessor was wrong, i.e., $\tilde t \ne t^*$, then we will, with high probability, observe a different outcome:
\begin{equation*}
    \overunderbraces{&&\br{3}{\text{By definition of an \emph{unstable example pair}}}}%
    {&f(t^*(u_0'))=&f(t^*(z(u_0)))&=&f(t^*(z(u_1)))&=f(t^*(u_1'))}%
    {&\br{2}{\text{By construction}}& &\br{2}{\text{By construction}}}
\end{equation*}
where the middle inequality holds true because
the examples $u_0$ and $u_1$ are an unstable example pair, and $z$ is the non-identity transformation used to construct $u_i'$ from $u_i$.

By coming up with multiple pre-images, querying the target pipeline, and observing the predictions, we can check whether our guess on the preprocessor is correct or not.

\subsubsection{A Greedy Pre-image Attack}

The previous step requires the ability to construct the pre-images for an arbitrary image $x$ and an arbitrary guessed transformation $\tilde t$.
While in general, this problem is intractable (e.g., a cryptographic hash function resists exactly this), common image preprocessors are not explicitly designed to be robust and so in practice, it is often nearly trivial.

In practice, we implement this attack via a greedy and naive attack that works
well for any transformation that operates over discrete integers
$t : \mathbb{Z}^n \to \mathbb{Z}^m$, which is the case for image preprocessors where pixel values lie between $0$ and $255$.

To begin, let $a_0$ be the image whose pre-image we would like to compute.
We then make random pixel-level perturbations to the image $a_0$ by randomly choosing a pixel coordinate $j$ and either increasing or decreasing its value by $1/255$.
We refer to each of these as $\{a_0^j\}_{j=0}^J$.
We take each of these candidates $a_0^j$ and check if $\tilde t(a_0^j) = \tilde t(a_0)$.
If any hold true, then we \emph{accept} this change and let $a_1 = a_0^j$.
We then repeat this procedure with $a_1$ to get a sequence of images $a_0, a_1 \dots a_K$ so that $\tilde t(a_0) = \dots = \tilde t(a_K)$ and that
$\lVert a_0 - a_K \rVert$ is sufficiently large.
We desire large perturbation because, intuitively, the larger the difference, the higher the probability that the unstable property will hold.
In other words, it is more likely that $f(t(z(u'_0))) = f(t(z(u'_1)))$ if $\tilde t \ne t$, where $u'_0$ and $u'_1$ are $a_K$ and $b_K$ in this case.
In practice, we only use one unstable example pair, but if more confidence is desired, an attacker could use many (at an increased query cost).
\section{Additional Experiment Results} \label{ap:sec:result}

\subsection{Complete Preprocessor-Aware Attack Results} \label{ap:ssec:atk_complete}

\begin{table*}[t]
    \small
    \caption{Comparing the mean adversarial perturbation norm ($\downarrow$) for cropping. The numbers in the parentheses indicate $s_o$ and $s_m$, respectively. ``$\Delta$'' is a ratio between the perturbation norm under a preprocessor-unaware (``Unaware'') vs our Bypassing Attack, both using their respectively best set of hyperparameters. The smallest adversarial distance found with untargeted and targeted attacks is in bold. For the distance, lower is better.}
    \label{tab:crop_app}
    \centering
    \begin{tabular}{@{}lllrrrrrrr@{}}
        \toprule
        \multirowcell{2}[0ex][l]{Preprocessors} & \multirowcell{2}[0ex][l]{Methods}      &
        \multirowcell{2}[0ex][l]{Hparams}       & \multicolumn{3}{c}{Untargeted Attacks} & \multicolumn{4}{c}{Targeted Attacks}                                                                                                                                                                                                                                                                                         \\ \cmidrule(lr){4-6} \cmidrule(lr){7-10}
                                                &                                        &                                      & Boundary                              & Sign-OPT                              & HSJA                                  & Boundary                              & Sign-OPT                              & HSJA                                  & QEBA                                  \\ \midrule
        \multirowcell{5}[0ex][l]{Crop                                                                                                                                                                                                                                                                                                                                                                                   \\($256 \to 224$)} & \multirowcell{2}[0ex][l]{Unaware} & Default & 11.1 & 6.7 & 4.4 & 48.6 & 50.6 & 40.9 & 24.7 \\
                                                &                                        & Best                                 & 5.3                                   & 6.5                                   & 4.2                                   & 42.8                                  & 50.4                                  & 38.2                                  & 22.2                                  \\ \cmidrule(lr){3-10}
                                                & \multirowcell{2}[0ex][l]{Bypassing                                                                                                                                                                                                                                                                                                                                    \\(ours)} & Default & 9.6 & 5.9 & 3.9 & 42.3 & 46.0 & 35.1 & 21.2 \\
                                                &                                        & Best                                 & 4.6                                   & 5.8                                   & \textbf{3.6}                          & 37.3                                  & 46.0                                  & 32.9                                  & \textbf{19.6}                         \\
                                                &                                        & $\Delta$                             & \textcolor{ForestGreen}{$1.16\times$} & \textcolor{ForestGreen}{$1.12\times$} & \textcolor{ForestGreen}{$1.16\times$} & \textcolor{ForestGreen}{$1.15\times$} & \textcolor{ForestGreen}{$1.08\times$} & \textcolor{ForestGreen}{$1.16\times$} & \textcolor{ForestGreen}{$1.13\times$} \\
        \bottomrule
    \end{tabular}
\end{table*}

\begin{table*}[t]
    \small
    \caption{Comparing the mean adversarial perturbation norm for resizing with different interpolation methods and input sizes. We report only the best hyperparameter found after the sweep. Lower is better.}
    \label{tab:resize_all_app}
    \centering
    \begin{tabular}{@{}llrrrrrrr@{}}
        \toprule
        \multirowcell{2}[0ex][l]{Preprocessors} & \multirowcell{2}[0ex][l]{Methods}                                                                                                                                                                                                                                                                                           & \multicolumn{3}{c}{Untargeted Attacks} & \multicolumn{4}{c}{Targeted Attacks}                                                              \\ \cmidrule(lr){3-5} \cmidrule(lr){6-9}                                  &                                        & Boundary                              & Sign-OPT                              & HSJA                                  & Boundary                              & Sign-OPT                              & HSJA          & QEBA \\ \midrule
        \multirowcell{4}[0ex][l]{Resize                                                                                                                                                                                                                                                                                                                                                                                                                                                                                    \\($1024 \to 224$)\\(Nearest)} & Unaware & 21.2                                  & 24.8                                  & 16.5                                  & 172.2                                 & 198.8                                 & 153.4                                 & 90.5          \\
                                                & SNS                                                                                                                                                                                                                                                                                                                         & n/a                                    & n/a                                  & 3.9          & n/a  & n/a  & 112.6         & 32.2          \\
                                                & Bypassing (ours)                                                                                                                                                                                                                                                                                                            & 4.7                                    & 5.8                                  & \textbf{3.7} & 37.7 & 46.3 & 33.3          & \textbf{19.4} \\
                                                & Biased-Grad (ours)                                                                                                                                                                                                                                                                                                          & n/a                                    & n/a                                  & 3.6          & n/a  & n/a  & 32.9          & 19.6          \\
        \cmidrule(l){2-9}
        \multirowcell{4}[0ex][l]{Resize                                                                                                                                                                                                                                                                                                                                                                                                                                                                                    \\($512 \to 224$)\\(Nearest)} & Unaware & 10.3                                  & 12.5                                  & 8.1                                   & 84.7                                  & 97.8                                  & 74.2                                  & 44.5          \\
                                                & SNS                                                                                                                                                                                                                                                                                                                         & n/a                                    & n/a                                  & 3.7          & n/a  & n/a  & 56.5          & 54.1          \\
                                                & Bypassing (ours)                                                                                                                                                                                                                                                                                                            & 4.5                                    & 5.7                                  & \textbf{3.6} & 37.3 & 45.5 & 32.6          & \textbf{19.4} \\
                                                & Biased-Grad (ours)                                                                                                                                                                                                                                                                                                          & n/a                                    & n/a                                  & 3.6          & n/a  & n/a  & 34.2          & 19.9          \\
        \cmidrule(l){2-9}
        \multirowcell{4}[0ex][l]{Resize                                                                                                                                                                                                                                                                                                                                                                                                                                                                                    \\($256 \to 224$)\\(Nearest)} & Unaware & 6.3                                   & 6.1                                   & 3.9                                   & 41.0                                  & 50.6                                  & 36.1                                  & 20.1          \\
                                                & SNS                                                                                                                                                                                                                                                                                                                         & n/a                                    & n/a                                  & 3.4          & n/a  & n/a  &            34.5   & 30.0              \\
                                                & Bypassing (ours)                                                                                                                                                                                                                                                                                                            & 7.7                                    & 5.4                                  & 3.4          & 36.0 & 44.8 & 31.3          & 17.9 \\
                                                & Biased-Grad (ours)                                                                                                                                                                                                                                                                                                          & n/a                                    & n/a                                  & \textbf{3.4} & n/a  & n/a  & 31.4          & \textbf{17.6}          \\ \cmidrule(l){2-9}
        \multirowcell{4}[0ex][l]{Resize                                                                                                                                                                                                                                                                                                                                                                                                                                                                                    \\($1024 \to 224$)\\(Bilinear)} & Unaware & 32.7                                  & 38.2                                  & 25.5                                  & 198.3                                 & 213.0                                 & 188.4                                 & 90.3          \\
                                                & SNS                                                                                                                                                                                                                                                                                                                         & n/a                                    & n/a                                  &      5.7        & n/a  & n/a  & 113.7         & 111.6         \\
                                                & Bypassing                                                                                                                                                                                                                                                                                                            (ours) & 7.4                                    & 9.1                                  & 6.0 & 58.2 & 70.9 & 50.3          & \textbf{30.0} \\
                                                & Biased-Grad (ours)                                                                                                                                                                                                                                                                                                          & n/a                                    & n/a                                  & \textbf{5.6}          & n/a  & n/a  & 56.4          & 36.4          \\
        \cmidrule(l){2-9}
        \multirowcell{4}[0ex][l]{Resize                                                                                                                                                                                                                                                                                                                                                                                                                                                                                    \\($512 \to 224$)\\(Bilinear)} & Unaware & 15.9                                  & 19.1                                  & 12.6                                  & 98.7                                  & 106.0                                 & 90.8                                  & 45.6          \\
                                                & SNS                                                                                                                                                                                                                                                                                                                         & n/a                                    & n/a                                  & \textbf{5.5} & n/a  & n/a  & 65.1          & 57.0          \\
                                                & Bypassing (ours)                                                                                                                                                                                                                                                                                                            & 7.4                                    & 9.2                                  & 5.9          & 57.7 & 70.9 & 50.2          & \textbf{30.3} \\
                                                & Biased-Grad (ours)                                                                                                                                                                                                                                                                                                          & n/a                                    & n/a                                  & 5.5          & n/a  & n/a  & 51.1          & 30.5          \\
        \cmidrule(l){2-9}
        \multirowcell{4}[0ex][l]{Resize                                                                                                                                                                                                                                                                                                                                                                                                                                                                                    \\($256 \to 224$)\\(Bilinear)} & Unaware & 6.3                                   & 7.8                                   & 5.1                          & 45.6                                  & 53.0                                  & 40.8                                  & 21.9          \\
                                                & SNS                                                                                                                                                                                                                                                                                                                         & n/a                                    & n/a                                  & 3.6          & n/a  & n/a  & 33.9 & 29.0          \\
                                                & Bypassing (ours)                                                                                                                                                                                                                                                                                                            & 7.7                                    & 9.9                                  & 6.1          & 45.5 & 57.8 & 46.2          & \textbf{21.5} \\
                                                & Biased-Grad (ours)                                                                                                                                                                                                                                                                                                          & n/a                                    & n/a                                  & \textbf{3.4} & n/a  & n/a  & 34.0          & 22.0          \\
        \cmidrule(l){2-9}
        \multirowcell{4}[0ex][l]{Resize                                                                                                                                                                                                                                                                                                                                                                                                                                                                                    \\($1024 \to 224$)\\(Bicubic)} & Unaware & 25.7                                  & 29.2                                  & 20.6                                  & 184.8                                 & 207.3                                 & 171.6                                 & 91.2                                  \\
                                                & SNS                                                                                                                                                                                                                                                                                                                         & n/a                                    & n/a                                  & 4.5          & n/a  & n/a  &        108.3       &     55.7          \\
                                                & Bypassing (ours)                                                                                                                                                                                                                                                                                                            & 5.8                                    & 7.1                                  & \textbf{4.5} & 46.4 & 57.7 & 40.6          & \textbf{23.8} \\
                                                & Biased-Grad (ours)                                                                                                                                                                                                                                                                                                          & n/a                                    & n/a                                  & 4.6          & n/a  & n/a  & 45.8          & 28.1          \\
        \cmidrule(l){2-9}
        \multirowcell{4}[0ex][l]{Resize                                                                                                                                                                                                                                                                                                                                                                                                                                                                                    \\($512 \to 224$)\\(Bicubic)} & Unaware & 13.1                                  & 15.4                                  & 10.1                                  & 91.1                                  & 101.5                                 & 81.1                                  & 44.3                                  \\
                                                & SNS                                                                                                                                                                                                                                                                                                                         & n/a                                    & n/a                                  & 4.6          & n/a  & n/a  & 59.9          & 55.8          \\
                                                & Bypassing (ours)                                                                                                                                                                                                                                                                                                            & 5.8                                    & 7.0                                  & \textbf{4.5} & 46.4 & 56.6 & 40.2          & \textbf{24.4} \\
                                                & Biased-Grad (ours)                                                                                                                                                                                                                                                                                                          & n/a                                    & n/a                                  & 4.5          & n/a  & n/a  & 42.6          & 25.6          \\ \cmidrule(l){2-9}
        \multirowcell{4}[0ex][l]{Resize                                                                                                                                                                                                                                                                                                                                                                                                                                                                                    \\($256 \to 224$)\\(Bicubic)} & Unaware & 6.0                                   & 7.4                                   & 4.8                                   & 44.2                                  & 51.9                                  & 39.4                                  & 21.5                         \\
                                                & SNS                                                                                                                                                                                                                                                                                                                         & n/a                                    & n/a                                  & 4.0          & n/a  & n/a  & 74.9          & 87.6          \\
                                                & Bypassing (ours)                                                                                                                                                                                                                                                                                                            & 5.8                                    & 7.3                                  & 4.6          & 42.5 & 52.9 & 37.6          & 21.6          \\
                                                & Biased-Grad (ours)                                                                                                                                                                                                                                                                                                          & n/a                                    & n/a                                  & \textbf{3.9} & n/a  & n/a  & 36.4 &     \textbf{21.2}          \\
        \bottomrule
    \end{tabular}
\end{table*}

\begin{table}[t]
    \small
    \caption{Comparison of the mean adversarial perturbation norm ($\downarrow$) for JPEG compression among the baseline attack unaware of the preprocessor, SNS, and our \nbp Attack. ``$\Delta$'' is a ratio between the perturbation norm under a preprocessor-unaware (``Unaware'') vs our Bypassing Attack, both using their respectively best set of hyperparameters.}\label{tab:jpeg}
    \centering
    \begin{tabular}{@{}lllrrr@{}}
        \toprule
        \multirowcell{2}[0ex][l]{Preprocess}                                                                                                                                                                                                            & \multirowcell{2}[0ex][l]{Methods}                                                                                                                                                                   &
        \multirowcell{2}[0ex][l]{Attack Hyperparameter}                                                                                                                                                                                                 & \multicolumn{1}{c}{Untargeted}                                                                                                                                                                      & \multicolumn{2}{c}{Targeted}                                                                                                                         \\ \cmidrule(lr){4-4} \cmidrule(lr){5-6}
                                                                                                                                                                                                                                                        &                                                                                                                                                                                                     &                              & HSJA                                  & HSJA                                  & QEBA                                  \\ \midrule
        \multirowcell{7}[0ex][l]{JPEG                                                                                                                                                                                                    (quality 100)} & \multirowcell{2}[0ex][l]{Unaware}                                                                                                                                                                   & Default                      & 5.7                                   & 35.8                                  & 18.8                                  \\
                                                                                                                                                                                                                                                        &                                                                                                                                                                                                     & Best                         & 3.5                                   & 31.9                                  & \textbf{18.8}                         \\ \cmidrule(l){3-6}
                                                                                                                                                                                                                                                        &
        \multirowcell{2}[0ex][l]{SNS}                                                                                                                                                                                                                   & Default                                                                                                                                                                                             & 3.1                          & 43.0                                  & 24.7                                                                          \\
                                                                                                                                                                                                                                                        &                                                                                                                                                                                                     & Best                         & 2.4                                   & \textbf{29.5}                         & 24.7                                  \\ \cmidrule(l){3-6}
                                                                                                                                                                                                                                                        & \multirowcell{2}[0ex][l]{Biased-Grad (ours)}                                                                                                                                                        & Default                      & 28.9                                  & 71.9                                  & 19.2                                  \\
                                                                                                                                                                                                                                                        &                                                                                                                                                                                                     & Best                         & \textbf{2.8}                          & 32.5                                  & 19.2                                  \\
                                                                                                                                                                                                                                                        &                                                                                                                                                                                                     & $\Delta$                     & \textcolor{ForestGreen}{$1.23\times$} & \textcolor{red}{$0.98\times$}         & \textcolor{red}{$0.98\times$}         \\
        \cmidrule(l){2-6}
        \multirowcell{7}[0ex][l]{JPEG                                                                                                                                                                                                    (quality 80)}  & \multirowcell{2}[0ex][l]{Unaware}                                                                                                                                                                   & Default                      & 29.6                                  & 85.7                                  & 50.7                                  \\
                                                                                                                                                                                                                                                        &                                                                                                                                                                                                     & Best                         & 8.9                                   & 63.2                                  & 43.9                                  \\ \cmidrule(l){3-6}
                                                                                                                                                                                                                                                        &
        \multirowcell{2}[0ex][l]{SNS}                                                                                                                                                                                                                   & Default                                                                                                                                                                                             & 13.2                         & 64.3                                  & 47.2                                                                          \\
                                                                                                                                                                                                                                                        &                                                                                                                                                                                                     & Best                         & 7.3                                   & 45.4                                  & 47.2                                  \\ \cmidrule(l){3-6}
                                                                                                                                                                                                                                                        & \multirowcell{2}[0ex][l]{Biased-Grad                                                                                                                                                        (ours)} & Default                      & 23.7                                  & 80.4                                  & 25.5                                  \\
                                                                                                                                                                                                                                                        &                                                                                                                                                                                                     & Best                         & \textbf{2.3}                          & \textbf{29.2}                         & \textbf{21.9}                         \\
                                                                                                                                                                                                                                                        &                                                                                                                                                                                                     & $\Delta$                     & \textcolor{ForestGreen}{$3.87\times$} & \textcolor{ForestGreen}{$2.16\times$} & \textcolor{ForestGreen}{$2.00\times$} \\
        \cmidrule(l){2-6}
        \multirowcell{7}[0ex][l]{JPEG                                                                                                                                                                                                    (quality 60)}  & \multirowcell{2}[0ex][l]{Unaware}                                                                                                                                                                   & Default                      & 29.2                                  & 86.8                                  & 56.1                                  \\
                                                                                                                                                                                                                                                        &                                                                                                                                                                                                     & Best                         & 9.2                                   & 63.2                                  & 52.7                                  \\ \cmidrule(l){3-6}
                                                                                                                                                                                                                                                        &
        \multirowcell{2}[0ex][l]{SNS}                                                                                                                                                                                                                   & Default                                                                                                                                                                                             & 11.9                         & 66.1                                  & 44.6                                                                          \\
                                                                                                                                                                                                                                                        &                                                                                                                                                                                                     & Best                         & 2.7                                   & 44.5                                  & 44.6                                  \\ \cmidrule(l){3-6}
                                                                                                                                                                                                                                                        & \multirowcell{2}[0ex][l]{Biased-Grad                                                                                                                                                        (ours)} & Default                      & 22.2                                  & 82.0                                  & 27.0                                  \\
                                                                                                                                                                                                                                                        &                                                                                                                                                                                                     & Best                         & \textbf{1.5}                          & \textbf{25.1}                         & \textbf{26.1}                         \\
                                                                                                                                                                                                                                                        &                                                                                                                                                                                                     & $\Delta$                     & \textcolor{ForestGreen}{$6.13\times$} & \textcolor{ForestGreen}{$2.51\times$} & \textcolor{ForestGreen}{$2.02\times$} \\
        \bottomrule
    \end{tabular}
\end{table}
\begin{table}[t]
    \small
    \caption{Comparison of the mean adversarial perturbation norm ($\downarrow$) for quantization among the baseline attack unaware of the preprocessor, SNS, and our \nbp Attack. ``$\Delta$'' is a ratio between the perturbation norm under a preprocessor-unaware (``Unaware'') vs our Bypassing Attack, both using their respectively best set of hyperparameters.}\label{tab:quantize}
    \centering
    \small
    \begin{tabular}{@{}lllrrr@{}}
        \toprule
        \multirowcell{2}[0ex][l]{Preprocess}            & \multirowcell{2}[0ex][l]{Methods}            &
        \multirowcell{2}[0ex][l]{Attack Hyperparameter} & \multicolumn{1}{c}{Untargeted}               & \multicolumn{2}{c}{Targeted}                                                                                                                         \\ \cmidrule(lr){4-4} \cmidrule(lr){5-6}
                                                        &                                              &                              & HSJA                                  & HSJA                                  & QEBA                                  \\ \midrule
        \multirowcell{7}[0ex][l]{Quantize (8 bits)}     & \multirowcell{2}[0ex][l]{Unaware}            & Default                      & 29.1                                  & 83.6                                  & 26.5                                  \\
                                                        &                                              & Best                         & 5.0                                   & 45.6                                  & 26.5                                  \\ \cmidrule(l){3-6}
                                                        & \multirowcell{2}[0ex][l]{SNS}                & Default                      & 6.8                                   & 42.8                                  & 35.0                                  \\
                                                        &                                              & Best                         & 4.6                                   & 42.8                                  & 35.0                                  \\ \cmidrule(l){3-6}
                                                        & \multirowcell{2}[0ex][l]{Biased-Grad (ours)} & Default                      & 7.1                                   & 46.2                                  & 21.3                                  \\
                                                        &                                              & Best                         & \textbf{3.9}                          & \textbf{33.9}                         & \textbf{20.6}                         \\
                                                        &                                              & $\Delta$                     & \textcolor{ForestGreen}{$1.27\times$} & \textcolor{ForestGreen}{$1.35\times$} & \textcolor{ForestGreen}{$1.29\times$} \\
        \cmidrule(l){2-6}
        \multirowcell{7}[0ex][l]{Quantize (6 bits)}     & \multirowcell{2}[0ex][l]{Unaware}            & Default                      & 30.4                                  & 86.1                                  & 40.6                                  \\
                                                        &                                              & Best                         & 7.5                                   & 48.2                                  & 39.4                                  \\ \cmidrule(l){3-6}
                                                        & \multirowcell{2}[0ex][l]{SNS}                & Default                      & 17.5                                  & 76.2                                  & 43.7                                  \\
                                                        &                                              & Best                         & 5.9                                   & 46.8                                  & 43.7                                  \\ \cmidrule(l){3-6}
                                                        & \multirowcell{2}[0ex][l]{Biased-Grad (ours)} & Default                      & 11.1                                  & 56.7                                  & 25.1                                  \\
                                                        &                                              & Best                         & \textbf{3.9}                          & \textbf{34.2}                         & \textbf{23.3}                         \\
                                                        &                                              & $\Delta$                     & \textcolor{ForestGreen}{$1.92\times$} & \textcolor{ForestGreen}{$1.41\times$} & \textcolor{ForestGreen}{$1.69\times$} \\
        \cmidrule(l){2-6}
        \multirowcell{7}[0ex][l]{Quantize (4 bits)}     & \multirowcell{2}[0ex][l]{Unaware}            & Default                      & 32.3                                  & 88.9                                  & 58.4                                  \\
                                                        &                                              & Best                         & 9.7                                   & 63.7                                  & 56.4                                  \\ \cmidrule(l){3-6}
                                                        & \multirowcell{2}[0ex][l]{SNS}                & Default                      & 22.2                                  & 76.5                                  & 57.2                                  \\
                                                        &                                              & Best                         & 6.4                                   & 55.9                                  & 57.2                                  \\ \cmidrule(l){3-6}
                                                        & \multirowcell{2}[0ex][l]{Biased-Grad (ours)} & Default                      & 19.2                                  & 74.7                                  & 31.8                                  \\
                                                        &                                              & Best                         & \textbf{3.1}                          & \textbf{39.3}                         & \textbf{28.8}                         \\
                                                        &                                              & $\Delta$                     & \textcolor{ForestGreen}{$3.05\times$} & \textcolor{ForestGreen}{$1.54\times$} & \textcolor{ForestGreen}{$1.86\times$} \\
        \bottomrule
    \end{tabular}
\end{table}

Notice that, for the nearest resizing, our \bp Attack finds adversarial examples with about the same mean adversarial distance as the no-preprocessor case \textbf{regardless of the input dimension} (see \cref{tab:resize_all_app}).
It may seem counter-intuitive: one might expect that the $\ell_2$-norm of the adversarial perturbation scales with the square root of the input dimension.
This may be the case if a new classifier were trained on each of the different input sizes~\citep{shafahi_are_2019}.
However, this observation matches the intuition that similar to cropping, the nearest resize operation only keeps 224$\times$224 pixels regardless of the input dimension.
Hence, only the perturbation on these pixels matters to the prediction.

To build some more intuition on this phenomenon, let's consider a toy example of a binary classifier that simply classifies one-dimensional data, e.g., white and black pixels with values of 0 and 1 respectively, by using a 0.5 threshold.
To push a white pixel over the decision boundary (or the threshold, in this case) requires a perturbation of size 0.5.
Now consider a new set of inputs with size $2 \times 2$ and a nearest resize that maps the 2$\times$2 inputs to one pixel.
The classifier remains unchanged.
In this case, the nearest resize simply picks one pixel (say, the top left) out of the four pixels.
Which pixel is picked depends on the exact implementation but does not matter for our purpose here.
To attack this classifier from a 2$\times$2 input, the adversary still needs to change only the top left pixel by 0.5, and thus, the adversarial distance remains unchanged.
Even for larger input sizes, only one pixel will still be selected.
While this toy example explains resizing with the nearest interpolation, it does not necessarily apply to bilinear or bicubic.
Nonetheless, all of our experimental results support this hypothesis.

\begin{figure*}
  \centering
  \subfloat[No Preprocessor\label{fig:no_prep_hyp}]{%
    \includegraphics[width=0.3\textwidth]{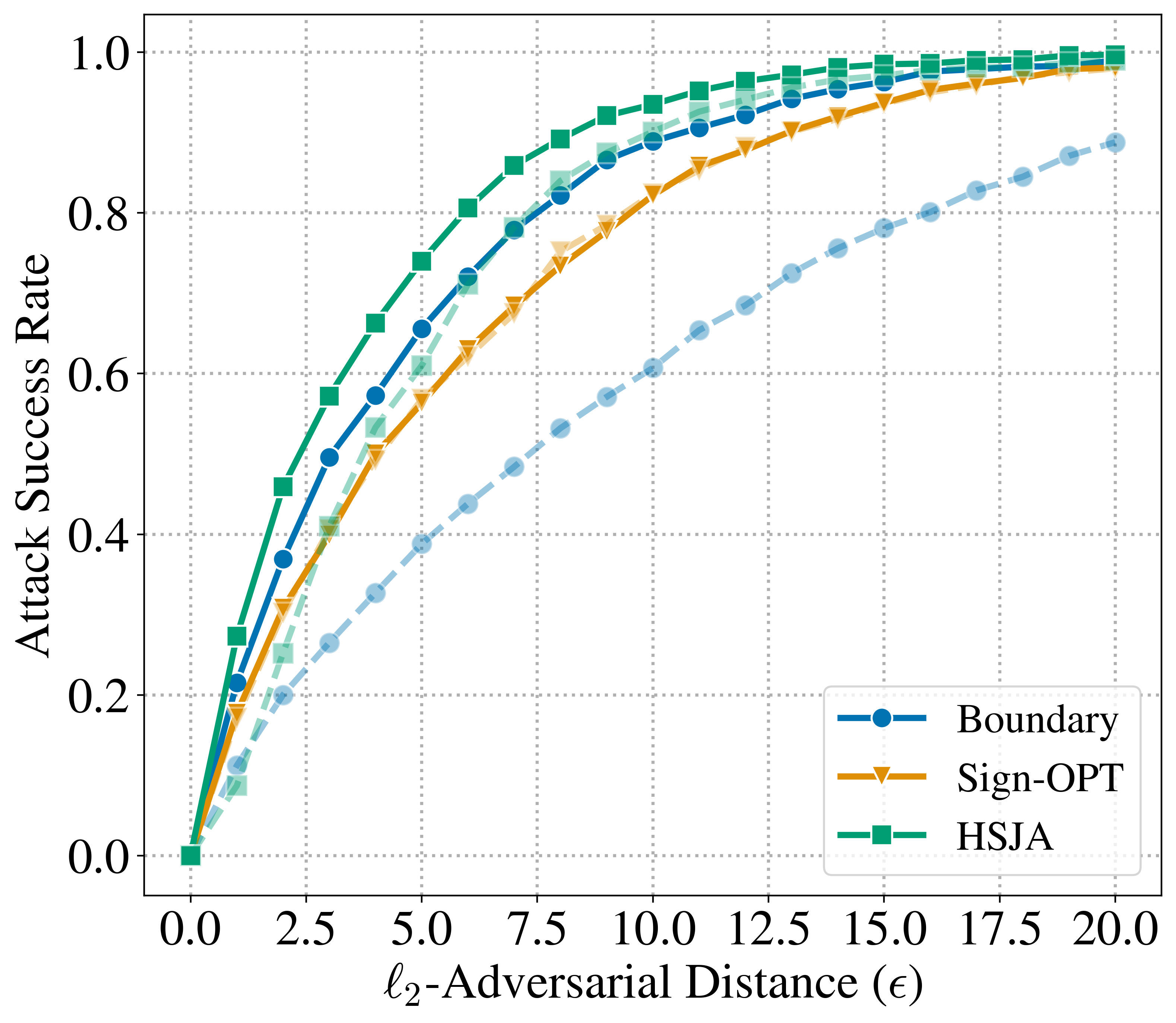}
  }
  \hfill
  \subfloat[Resize ($1024 \to 224$, nearest)\label{fig:resize_hyp}]{%
    \includegraphics[width=0.3\textwidth]{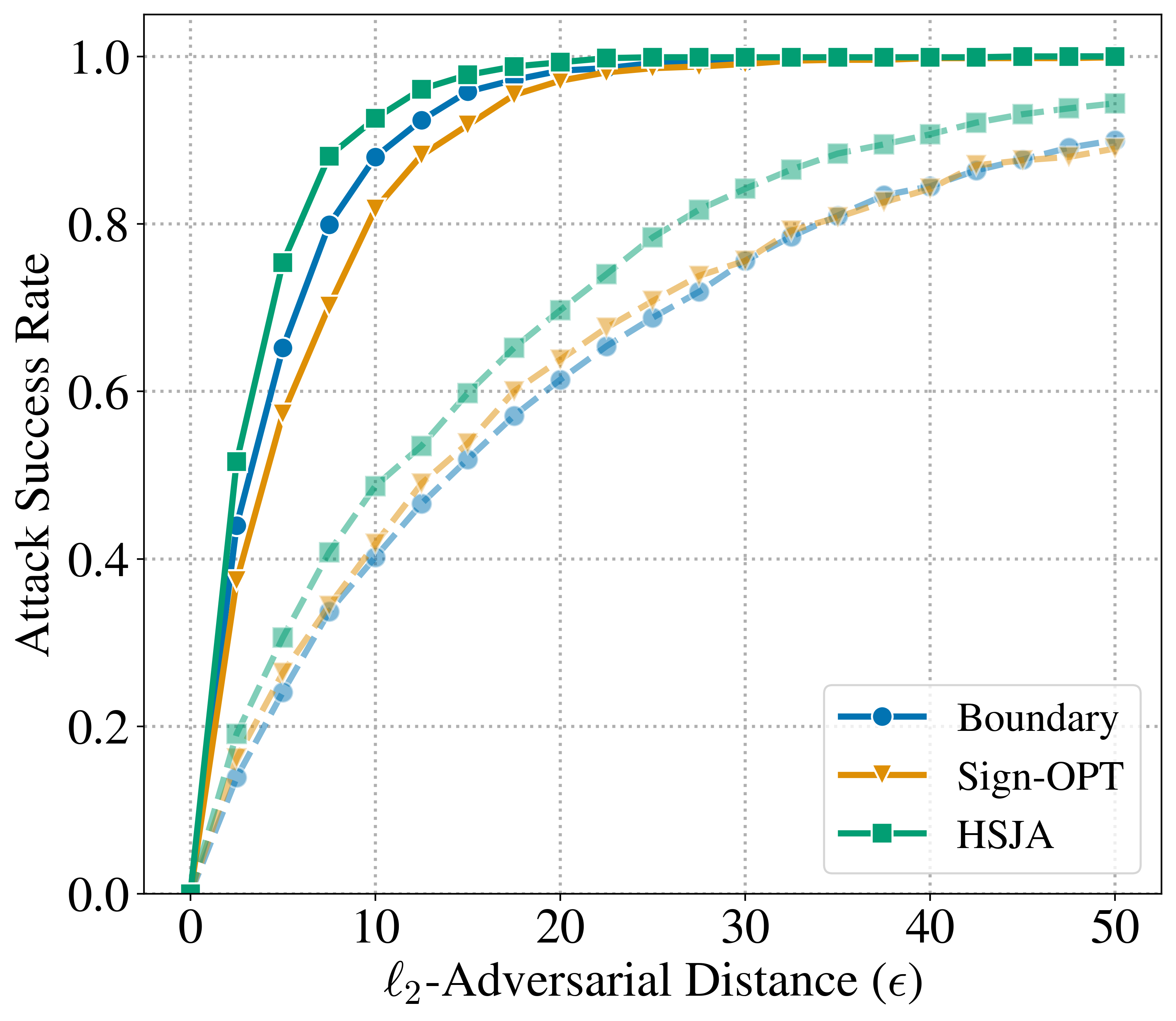}
  }
  \hfill
  \subfloat[Quantize (8 bits)\label{fig:quant_hyp}]{%
    \includegraphics[width=0.3\textwidth]{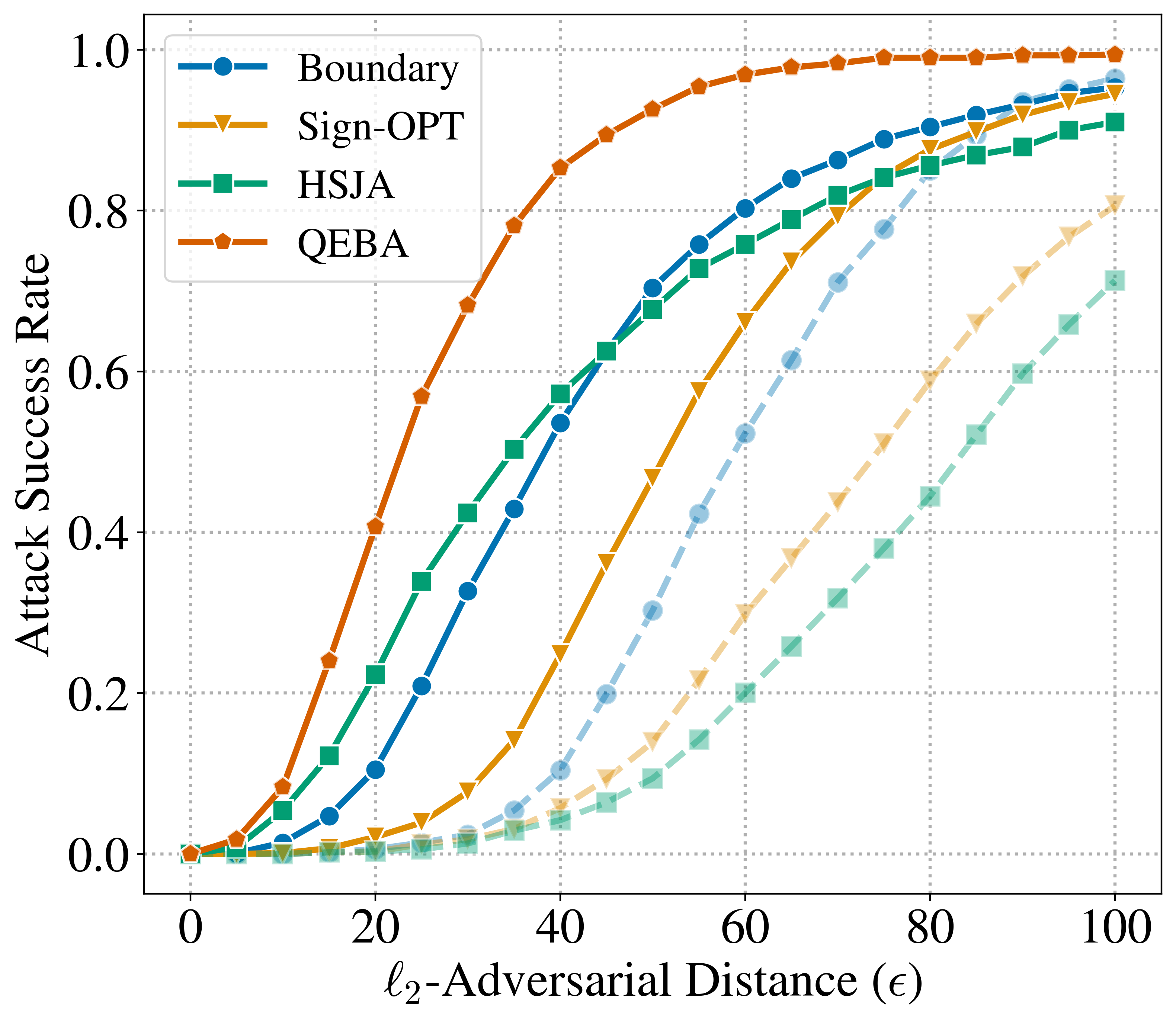}
  }
  \caption{Plots of the attack success rate at varying maximum adversarial distance with different preprocessors. The darker solid lines denote the preprocessor-unaware and the Bypassing attacks with their respectively best hyperparameters. The dashed lines denote the default hyperparameters, and the remaining lighter solid lines correspond to the other set of hyperparameters we sweep.}
  \label{fig:hyp_multi_prep}
\end{figure*}

\begin{figure*}
  \centering
  \subfloat[Boundary Attack\label{fig:boundary_hyp}]{%
    \includegraphics[width=0.3\textwidth]{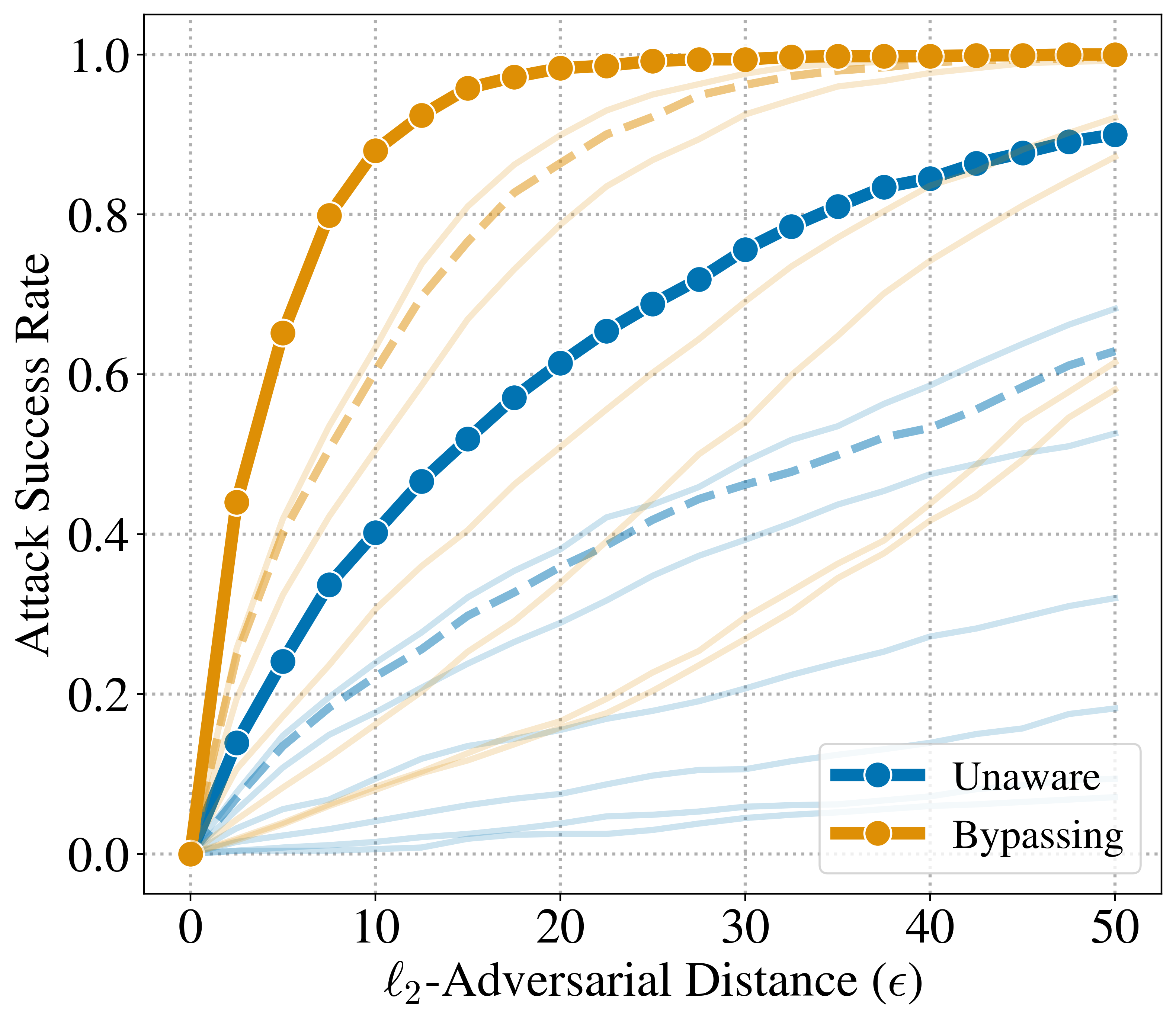}
  }
  \hfill
  \subfloat[Sign-OPT Attack\label{fig:signopt_hyp}]{%
    \includegraphics[width=0.3\textwidth]{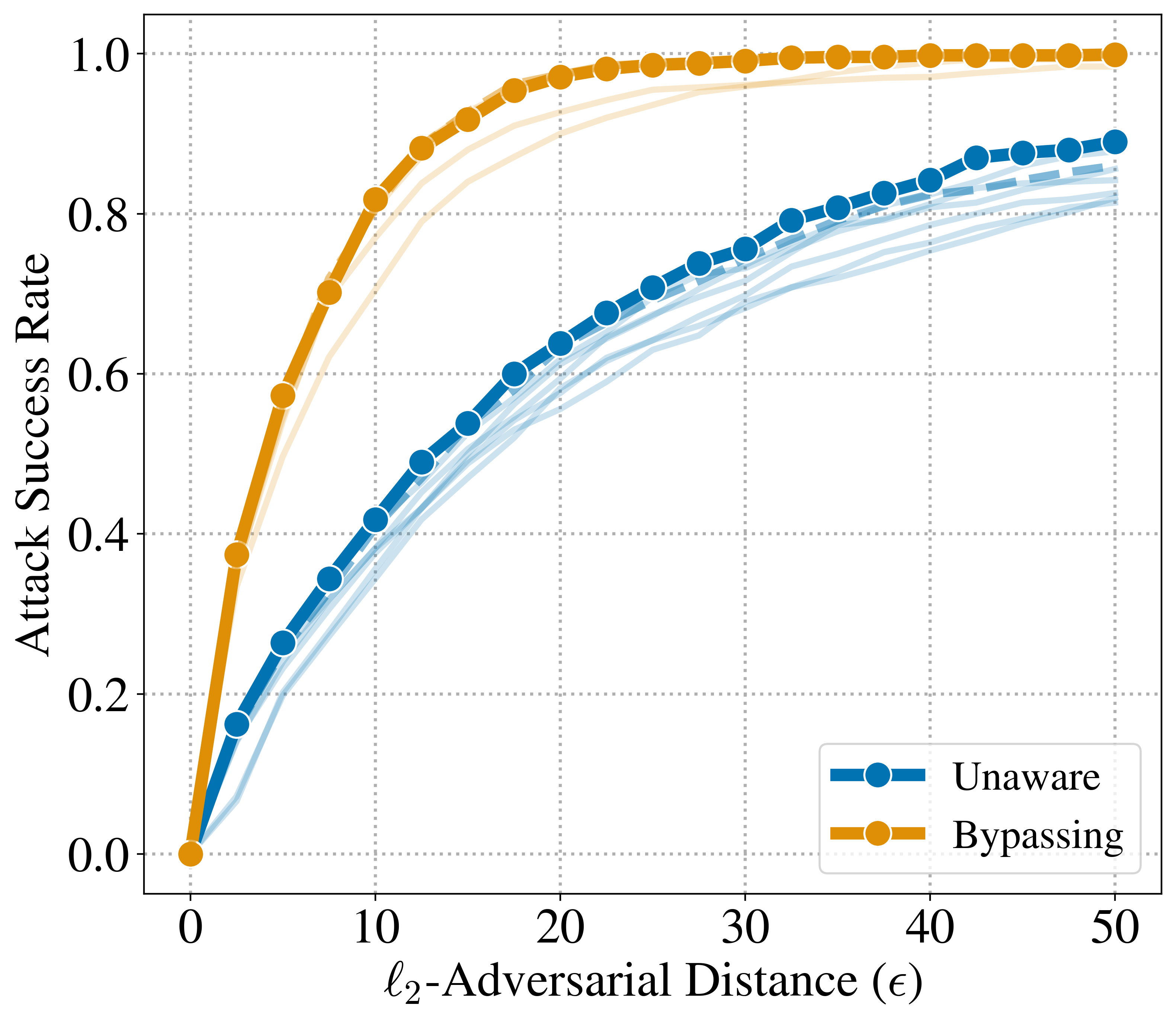}
  }
  \hfill
  \subfloat[HSJA\label{fig:hsja_hyp}]{%
    \includegraphics[width=0.3\textwidth]{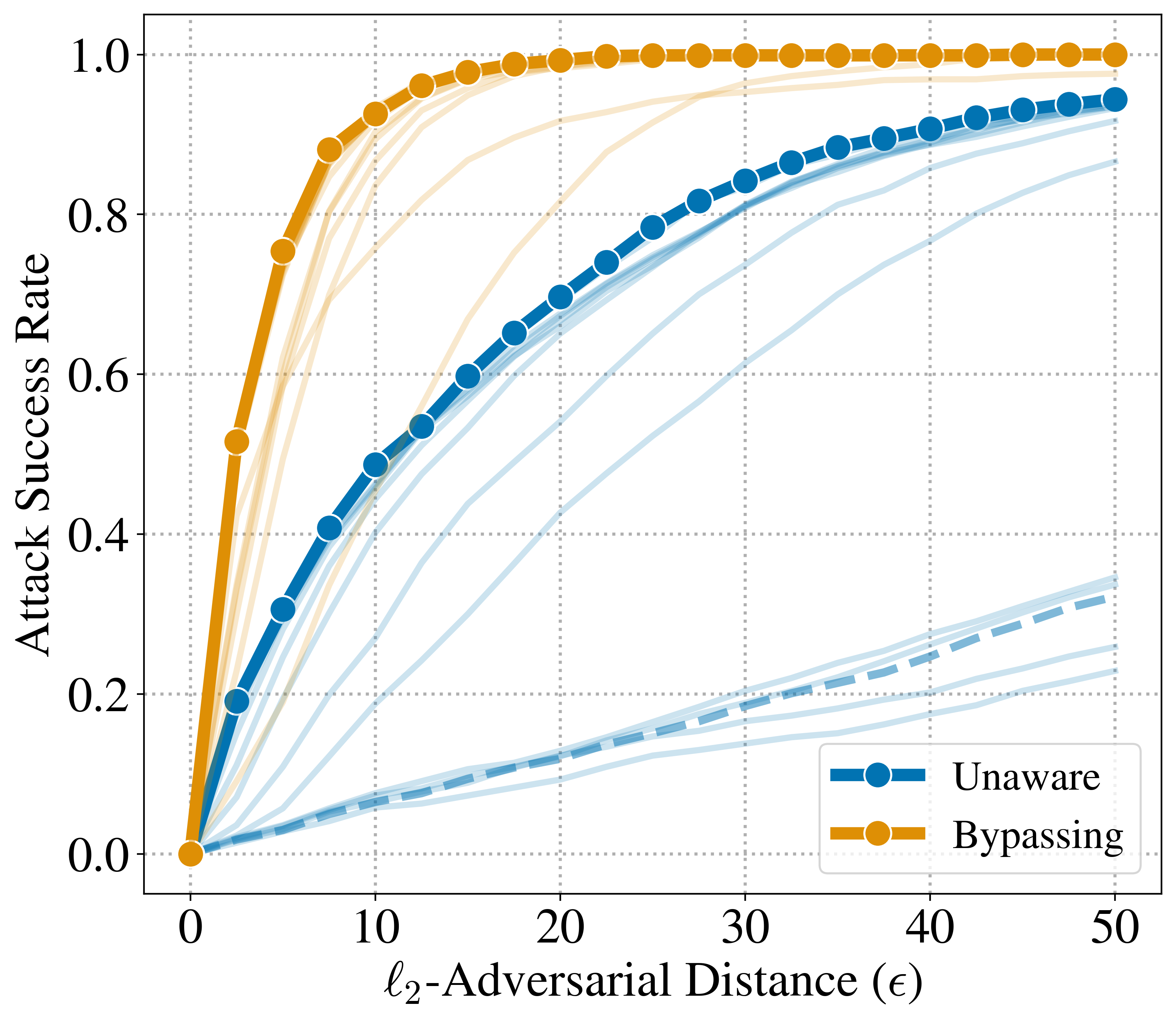}
  }
  \caption{Plots of the attack success rate at varying maximum adversarial distance. Here, the preprocessor is resizing with the nearest interpolation from $1024 \times 1024$ to $224 \times 224$, corresponding to the first five rows of \cref{tab:resize_all_app} (untargeted). The solid lines with markers denote the preprocessor-unaware and the \bp attacks with their respective best hyperparameters. The dashed lines denote the default hyperparameters, and the remaining lighter solid lines correspond to the other set of hyperparameters we sweep.}\label{fig:hyp_resize_utg}
\end{figure*}

\subsection{Attack Hyperparameter Choices}\label{ap:ssec:hyp}

We have seen from \cref{sec:attacks} that fine-tuning the hyperparameters improves the attack significantly in most cases.
We discuss when it is most important for the adversary to fine-tune their attack hyperparameters.
\cref{fig:hyp_resize_utg} (\cref{ap:sec:result}) shows the attack success rate at varying adversarial distances for three untargeted attack algorithms.
For Boundary, HSJA, and QEBA attacks, the gain from selecting the right set of hyperparameters is significant, a large improvement over the default.

For instance, a properly tuned Boundary attack outperforms Sign-OPT and HSJA attacks with their default hyperparameters in majority of the settings with resizing preprocessor.

For most attacks, we do not observe a universally good set of hyperparameters across different preprocessors.
However, there are two general rules of thumb when it comes to better guess the hyperparameters:
\begin{enumerate}
  \item Using a larger value of $\gamma$ ($10^3$--$10^4$) in HSJA attack is almost always better than the default ($10$). This applies to both preprocessor-aware and -unaware attacks and to all preprocessors.
  \item QEBA attack samples the noise used for gradient approximation from an image space with a smaller size $rs_o \times rs_o$ where $s_o$ is the original input size, and $r$ is the hyperparameter smaller than 1. The default value of $r$ is $\frac{1}{4}$ for $s_o=224$. Consequently, for a larger $s_o$ such as the resizing preprocessor, setting $r$ to be smaller accordingly is always beneficial. For example, we find that for $s_o=256,512,1024$, the best values of $r$ are $\frac{1}{8}$, $\frac{1}{16}$, $\frac{1}{32}$, respectively.
\end{enumerate}

Here, we include two figures that compare the effect of tuning the attack hyperparameters in multiple settings.
\cref{fig:hyp_multi_prep} suggests that the default hyperparameters often work well as expected when no preprocessor is used while there is much greater discrepancy between the default and the best hyperparameters when preprocessors are used.

The degree in which the hyperparameter tuning matters also depends on the attack algorithm.
\cref{fig:hyp_resize_utg} visually compares the effectiveness of three untargeted attacks on the resizing preprocessor.
It is obvious that Boundary and HSJA attacks benefit much more from a hyperparameter sweep compared to Sign-OPT attack.

\begin{table}[t]
    \small
    \caption{Comparison of the mean adversarial perturbation norm ($\downarrow$) for the resizing preprocessor with nearest interpolation (1024 $\to$ 224 pixels) among the baseline attack unaware of the preprocessor, SNS, and our \bp Attack under various hyperparameters of each attack algorithm. Notice the generally large gap between the best hyperparameter (\textbf{bold}) vs the default one (first row for each respective attack). The \underline{underlined number} is best for the given attack algorithm. $s_{\parallel}$ and $s_{\perp}$ denote the Boundary attack's step size in the parallel and the orthogonal direction to a given input, respectively. Note that SNS is only applicable to HSJA and QEBA attacks, and for QEBA, SNS replaces its subspace noise sampling process. Hence, we only report a single number for SNS + QEBA attack.}\label{tab:atk_params_resize}
    \vspace{-5pt}
    \centering
    \begin{tabular}{@{}lllrrr@{}}
        \toprule
        Preprocessor                     & Attack                & Attack Parameters                     & Unaware        & SNS            & Ours                      \\
        \midrule
        Resize (nearest, 1024 $\to$ 224) & Boundary (untargeted) & $s_{\parallel}=0.01, s_{\perp}=0.01$  & 45.4           &                & 9.8                       \\
                                         &                       & $s_{\parallel}=0.1, s_{\perp}=0.01$   & 56.1           &                & 12.4                      \\
                                         &                       & $s_{\parallel}=0.001, s_{\perp}=0.01$ & 134.1          &                & 29.1                      \\
                                         &                       & $s_{\parallel}=0.01, s_{\perp}=0.1$   & \textbf{21.2}  &                & \underline{\textbf{4.7}}  \\
                                         &                       & $s_{\parallel}=0.01, s_{\perp}=0.001$ & 40.6           &                & 8.8                       \\
        \cmidrule{3-6}
                                         & Sign-OPT (untargeted) & $\alpha=0.2, \beta=0.001$             & \textbf{24.8}  &                & 5.8                       \\
                                         &                       & $\alpha=2, \beta=0.001$               & 25.2           &                & 5.9                       \\
                                         &                       & $\alpha=0.02, \beta=0.001$            & 25.0           &                & \underline{\textbf{5.8}}  \\
                                         &                       & $\alpha=0.2, \beta=0.01$              & 25.3           &                & 5.8                       \\
                                         &                       & $\alpha=0.2, \beta=0.0001$            & 25.4           &                & 5.9                       \\
        \cmidrule{3-6}
                                         & HSJA (untargeted)     & $\gamma=10^1$                         & 28.5           &                & 3.8                       \\
                                         &                       & $\gamma=10^2$                         & 18.3           &                & \underline{\textbf{3.7}}  \\
                                         &                       & $\gamma=10^3$                         & 17.4           & 12.4           & 3.7                       \\
                                         &                       & $\gamma=10^4$                         & 16.8           & 4.8            & 3.8                       \\
                                         &                       & $\gamma=10^5$                         & \textbf{16.5}  & \textbf{3.9}   & 8.0                       \\
                                         &                       & $\gamma=10^6$                         & 26.2           & 6.1            & 60.6                      \\
        \cmidrule{3-6}
                                         & Boundary (targeted)   & $s_{\parallel}=0.01, s_{\perp}=0.01$  & 194.4          &                & 42.3                      \\
                                         &                       & $s_{\parallel}=0.1, s_{\perp}=0.01$   & 242.6          &                & 52.4                      \\
                                         &                       & $s_{\parallel}=0.001, s_{\perp}=0.01$ & 310.2          &                & 67.8                      \\
                                         &                       & $s_{\parallel}=0.01, s_{\perp}=0.1$   & 233.1          &                & 50.6                      \\
                                         &                       & $s_{\parallel}=0.01, s_{\perp}=0.001$ & \textbf{172.2} &                & \underline{\textbf{37.7}} \\
        \cmidrule{3-6}
                                         & Sign-OPT (targeted)   & $\alpha=0.2, \beta=0.001$             & 201.3          &                & \underline{\textbf{46.3}} \\
                                         &                       & $\alpha=2, \beta=0.001$               & \textbf{199.4} &                & 46.4                      \\
                                         &                       & $\alpha=0.02, \beta=0.001$            & 200.2          &                & 46.4                      \\
                                         &                       & $\alpha=0.2, \beta=0.01$              & 203.3          &                & 47.1                      \\
                                         &                       & $\alpha=0.2, \beta=0.0001$            & 202.4          &                & 46.4                      \\
        \cmidrule{3-6}
                                         & HSJA (targeted)       & $\gamma=10^1$                         & 168.3          &                & 35.2                      \\
                                         &                       & $\gamma=10^2$                         & 160.5          &                & 34.0                      \\
                                         &                       & $\gamma=10^3$                         & 159.7          & 122.7          & 33.3                      \\
                                         &                       & $\gamma=10^4$                         & \textbf{153.4} & \textbf{112.6} & \underline{\textbf{23.5}} \\
                                         &                       & $\gamma=10^5$                         & 162.0          & 212.5          & 37.3                      \\
        \cmidrule{3-6}
                                         & QEBA (targeted)       & Naive, $\gamma=0.01$                  & 138.7          & \textbf{32.2}  & 29.7                      \\
                                         &                       & Resize 2$\times$,$\gamma=0.01$        & 139.1          &                & 21.9                      \\
                                         &                       & Resize 4$\times$,$\gamma=0.01$        & 124.5          &                & \underline{\textbf{19.4}} \\
                                         &                       & Resize 8$\times$,$\gamma=0.01$        & 103.7          &                & 19.9                      \\
                                         &                       & Resize 16$\times$,$\gamma=0.01$       & 92.5           &                & 26.3                      \\
                                         &                       & Resize 32$\times$,$\gamma=0.01$       & \textbf{90.5}  &                & 42.9                      \\
        \bottomrule
    \end{tabular}
\end{table}

\begin{table}[t]
    \small
    \caption{Comparison of the mean adversarial perturbation norm ($\downarrow$) for the cropping preprocessor (256 $\to$ 224 pixels). For the full description, please refer to \cref{tab:atk_params_crop}.}\label{tab:atk_params_crop}
    \vspace{-5pt}
    \centering
    \begin{tabular}{@{}lllrrr@{}}
        \toprule
        Preprocessor         & Attack                & Attack Parameters                     & Unaware       & SNS           & Ours                      \\
        \midrule
        Crop (256 $\to$ 224) & Boundary (untargeted) & $s_{\parallel}=0.01, s_{\perp}=0.01$  & 11.1          &               & 9.6                       \\
                             &                       & $s_{\parallel}=0.1, s_{\perp}=0.01$   & \textbf{5.3}  &               & \underline{\textbf{4.6}}  \\
                             &                       & $s_{\parallel}=0.001, s_{\perp}=0.01$ & 32.8          &               & 28.5                      \\
                             &                       & $s_{\parallel}=0.01, s_{\perp}=0.1$   & 14.1          &               & 12.1                      \\
                             &                       & $s_{\parallel}=0.01, s_{\perp}=0.001$ & 10.1          &               & 8.7                       \\
        \cmidrule{3-6}
                             & Sign-OPT (untargeted) & $\alpha=0.2, \beta=0.001$             & 6.7           &               & 5.9                       \\
                             &                       & $\alpha=2, \beta=0.001$               & 6.8           &               & 5.9                       \\
                             &                       & $\alpha=0.02, \beta=0.001$            & 6.6           &               & \underline{\textbf{5.8}}  \\
                             &                       & $\alpha=0.2, \beta=0.01$              & \textbf{6.5}  &               & 5.9                       \\
                             &                       & $\alpha=0.2, \beta=0.0001$            & 6.7           &               & 5.9                       \\
        \cmidrule{3-6}
                             & HSJA (untargeted)     & $\gamma=10^2$                         & 4.4           & 3.7           & 3.6                       \\
                             &                       & $\gamma=10^3$                         & \textbf{4.2}  & \textbf{3.7}  & \underline{\textbf{3.6}}  \\
                             &                       & $\gamma=10^4$                         & 4.2           & 3.8           & 4.9                       \\
                             &                       & $\gamma=10^5$                         & 6.0           & 6.9           & 4.9                       \\
        \cmidrule{3-6}
                             & Boundary (targeted)   & $s_{\parallel}=0.01, s_{\perp}=0.01$  & 48.6          &               & 42.3                      \\
                             &                       & $s_{\parallel}=0.1, s_{\perp}=0.01$   & 62.0          &               & 53.1                      \\
                             &                       & $s_{\parallel}=0.001, s_{\perp}=0.01$ & 76.9          &               & 66.8                      \\
                             &                       & $s_{\parallel}=0.01, s_{\perp}=0.1$   & 58.0          &               & 50.5                      \\
                             &                       & $s_{\parallel}=0.01, s_{\perp}=0.001$ & \textbf{42.8} &               & \underline{\textbf{37.3}} \\
        \cmidrule{3-6}
                             & Sign-OPT (targeted)   & $\alpha=0.2, \beta=0.001$             & 52.8          &               & \underline{\textbf{46.3}} \\
                             &                       & $\alpha=2, \beta=0.001$               & \textbf{52.7} &               & 46.5                      \\
                             &                       & $\alpha=0.02, \beta=0.001$            & 52.9          &               & 46.4                      \\
                             &                       & $\alpha=0.2, \beta=0.01$              & 53.9          &               & 47.7                      \\
                             &                       & $\alpha=0.2, \beta=0.0001$            & 53.2          &               & 46.7                      \\
        \cmidrule{3-6}
                             & HSJA (targeted)       & $\gamma=10^1$                         & 40.4          &               & 34.9                      \\
                             &                       & $\gamma=10^2$                         & 38.7          & 36.4          & 33.6                      \\
                             &                       & $\gamma=10^3$                         & \textbf{38.2} & \textbf{35.4} & \underline{\textbf{32.9}} \\
                             &                       & $\gamma=10^4$                         & 47.4          & 46.3          & 44.7                      \\
                             &                       & $\gamma=10^5$                         & 104.5         & 104.2         & 92.9                      \\
        \cmidrule{3-6}
                             & QEBA (targeted)       & Naive, $\gamma=0.01$                  & 34.0          & \textbf{31.5} & 29.5                      \\
                             &                       & Resize 2$\times$,$\gamma=0.01$        & 24.7          &               & 21.2                      \\
                             &                       & Resize 4$\times$,$\gamma=0.01$        & \textbf{22.2} &               & \underline{\textbf{19.6}} \\
                             &                       & Resize 8$\times$,$\gamma=0.01$        & 23.2          &               & 20.3                      \\
                             &                       & Resize 16$\times$,$\gamma=0.01$       & 30.0          &               & 26.8                      \\
        \bottomrule
    \end{tabular}
\end{table}

\begin{table}[t]
    \small
    \caption{Comparison of the mean adversarial perturbation norm ($\downarrow$) for the quantization preprocessor (4 bits) and JPEG compression (quality of 60). For the full description, please refer to \cref{tab:atk_params_crop}.}\label{tab:atk_params_quant_jpeg}
    \vspace{-5pt}
    \centering
    \begin{tabular}{@{}lllrrr@{}}
        \toprule
        Preprocessor      & Attack            & Attack Parameters               & Unaware       & SNS           & Ours                      \\
        \midrule
        Quantize (4 bits) & HSJA (untargeted) & $\gamma=10^2$                   & 29.8          & 22.2          & 4.4                       \\
                          &                   & $\gamma=10^3$                   & 22.7          & 12.3          & \underline{\textbf{3.1}}  \\
                          &                   & $\gamma=10^4$                   & 11.2          & \textbf{6.4}  & 10.3                      \\
                          &                   & $\gamma=10^5$                   & \textbf{9.7}  & 123.3         & 42.6                      \\
        \cmidrule{3-6}
                          & HSJA (targeted)   & $\gamma=10^2$                   & 85.3          & 76.5          & 54.5                      \\
                          &                   & $\gamma=10^3$                   & 74.7          & 57.6          & \underline{\textbf{39.3}} \\
                          &                   & $\gamma=10^4$                   & \textbf{63.7} & \textbf{55.9} & 48.8                      \\
                          &                   & $\gamma=10^5$                   & 95.2          & 92.1          & 91.3                      \\
        \cmidrule{3-6}
                          & QEBA (targeted)   & Naive, $\gamma=0.01$            & 71.8          & \textbf{57.2} & 33.6                      \\
                          &                   & Resize 2$\times$,$\gamma=0.01$  & 61.2          &               & 31.7                      \\
                          &                   & Resize 4$\times$,$\gamma=0.01$  & 58.4          &               & 30.4                      \\
                          &                   & Resize 8$\times$,$\gamma=0.01$  & \textbf{56.4} &               & \underline{\textbf{28.8}} \\
                          &                   & Resize 16$\times$,$\gamma=0.01$ & 60.4          &               & 29.2                      \\
        \cmidrule{2-6}
        JPEG (quality 60) & HSJA (untargeted) & $\gamma=10^2$                   & 27.5          & 11.9          & 10.2                      \\
                          &                   & $\gamma=10^3$                   & 21.3          & 6.6           & 5.0                       \\
                          &                   & $\gamma=10^4$                   & 10.5          & \textbf{2.7}  & 2.0                       \\
                          &                   & $\gamma=10^5$                   & \textbf{9.2}  & 3.4           & \underline{\textbf{1.5}}  \\
        \cmidrule{3-6}
                          & HSJA (targeted)   & $\gamma=10^2$                   & 80.5          & 66.1          & 66.1                      \\
                          &                   & $\gamma=10^3$                   & 66.4          & \textbf{44.5} & 43.5                      \\
                          &                   & $\gamma=10^4$                   & \textbf{63.2} & 51.8          & \underline{\textbf{25.1}} \\
                          &                   & $\gamma=10^5$                   & 93.3          &               & 93.5                      \\
        \cmidrule{3-6}
                          & QEBA (targeted)   & Naive, $\gamma=0.01$            & 64.9          & \textbf{44.6} & 28.3                      \\
                          &                   & Resize 2$\times$,$\gamma=0.01$  & 58.5          &               & 21.1                      \\
                          &                   & Resize 4$\times$,$\gamma=0.01$  & 56.1          &               & \underline{\textbf{21.0}} \\
                          &                   & Resize 8$\times$,$\gamma=0.01$  & \textbf{52.7} &               & 22.7                      \\
                          &                   & Resize 16$\times$,$\gamma=0.01$ & 53.3          &               & 25.9                      \\
        \bottomrule
    \end{tabular}
\end{table}

For the numerical comparison between the best and the default hyperparameter choices on quantization and JPEG, please refer to \cref{tab:quantize} and \cref{tab:jpeg}, respectively.
\cref{tab:atk_params_resize,tab:atk_params_crop,tab:atk_params_quant_jpeg} show the results with all the hyperparameters we sweep for resizing, cropping, and quantization/JPEG, respectively.
In the most extreme case, the best hyperparameters reduce the mean adversarial distance by a factor of 15$\times$ for JPEG with quality 60 under untargeted HSJA + our Biased-Gradient attack.

\subsection{Varying Number of Attack Iterations}\label{ap:ssec:num_steps}

There are two interesting properties we observe when we vary the number of queries the adversary can utilize.
So far we have considered attacks that use exactly 5,000 queries;
in this section, we now test attacks with 500 to 50,000 queries.
\cref{fig:num_steps} plots the mean adversarial distance as a function of the number of queries for QEBA attack with the best hyperparameter for each respective setting.
First, the adversarial distance plateaus after around 10,000 queries, and the distance found by preprocessor-unaware attacks never reaches that of Bypassing/\nbp Attack.
This suggests that our preprocessor-aware attack does not only improve the efficiency of the attack algorithms but also allows it to find closer adversarial examples that would have been completely missed otherwise.

The second observation is that the improvement from \bp Attack over the preprocessor-unaware attack is consistent across all numbers of queries.
For instance, in \cref{fig:num_steps_resize}, the Bypassing Attack reduces the mean adversarial distance by a factor of around 4.5 to 4.8 for any number of queries.
This is not the case for the \nbp Attack which is relatively more effective at a larger number of queries.
In \cref{fig:num_steps_jpeg}, \nbp Attack yields an improvement of 1.1$\times$ at 500 queries and 2.5$\times$ beyond 10,000 queries.

\begin{figure}[t]
  \centering
  \includegraphics[width=0.7\textwidth]{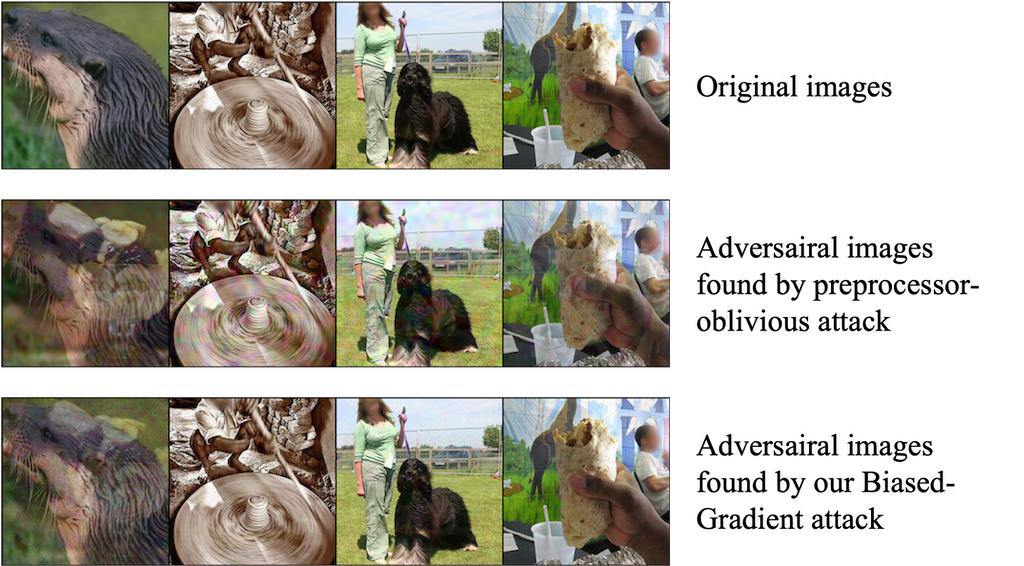}
  \caption{Randomly selected test images including the original and the adversarial. Our \nbp attack finds adversarial examples with a noticeably less perceptible perturbation compared to the ones generated by the preprocessor-unaware baseline. The target preprocessor here is 8-bit quantization, and the base attack is targeted QEBA. Looking closely, one can see some detail of the initialized images from the target class left in the adversarial images.}\label{fig:img_samples}
\end{figure}

\end{document}